\documentclass[fleqn,usenatbib,useAMS]{mnras}

\usepackage{xspace} 

\usepackage{graphicx}	
\usepackage{amsmath}	
\usepackage{amssymb}	
\usepackage{multicol}        
\usepackage{bm}		
\usepackage{pdflscape}	

\renewcommand{\d}{\mathrm{d}}
\newcommand{\Mpc}{\ensuremath{\textrm{Mpc}}}

\newcommand{\Cl}{C_{\ell}}
\newcommand{\Clp}{C_{\ell'}}

\newcommand{\lmax}{\ell_{\mathrm{max}}}

\newcommand{\fsky}{f_{\mathrm{sky}}}

\newcommand{\As}{\ensuremath{A_\textrm{s}}}
\newcommand{\sigmaeight}{\sigma_{8}}
\newcommand{\Seight}{S_{8}}
\newcommand{\Omegac}{\Omega_{\mathrm{c}}}

\newcommand{\Omegam}{\Omega_{\mathrm{m}}}

\newcommand{\Mb}{M_{\textrm{b}}}
\newcommand{\Tagn}{T_{\textrm{AGN}}}

\newcommand{\Halofit}{\textsc{HaloFit}\xspace}
\newcommand{\HMCode}{\textsc{HMCode}\xspace}
\newcommand{\Bacco}{\textsc{Bacco}\xspace}
\newcommand{\BCEmu}{\textsc{BCemu}\xspace}
\newcommand{\Cosmosis}{\texttt{CosmoSiS}\xspace}
\newcommand{\KiDS}{\textrm{KiDS-1000}\xspace}
\newcommand{\DES}{\textrm{DES-Y3}\xspace}

\newcommand{\Bahamas}{\textsc{Bahamas}\xspace}
\newcommand{\Eagle}{\textsc{Eagle}\xspace}
\newcommand{\HZAGN}{\textsc{Horizon-AGN}\xspace}
\newcommand{\Illustris}{\textsc{Illustris}\xspace}
\newcommand{\Simba}{\textsc{Simba}\xspace}
\newcommand{\TNG}{\textsc{Illustris-TNG}\xspace}
\newcommand{\Camels}{\textsc{Camels}\xspace}

\newcommand{\Phy}{\ensuremath{P^{\textsc{Hy}}(k)}}
\newcommand{\Pdm}{\ensuremath{P^{\textsc{DM}}(k)}}

\newcommand{\Phykz}{\ensuremath{P^{\textsc{Hy}}(k, z)}}
\newcommand{\Pdmkz}{\ensuremath{P^{\textsc{DM}}(k, z)}}

\usepackage[T1]{fontenc}
\usepackage{ae,aecompl}
\usepackage{newtxtext,newtxmath}

\title[Theoretical uncertainties for cosmic shear]{
  Mitigating baryon feedback bias in cosmic shear through a 
  theoretical error covariance in the matter power spectrum
  }

\author[A. Maraio et al.]{Alessandro Maraio%
\thanks{Contact e-mail: \href{mailto:maraio@roe.ac.uk}{maraio@roe.ac.uk}},
Alex Hall, Andy Taylor%
\\
Institute for Astronomy, University of Edinburgh, Royal Observatory, Blackford Hill, Edinburgh EH9 3HJ, UK}

\date{Last updated 16 October 2024}

\pubyear{2024}

\begin{document}
\label{firstpage}
\pagerange{\pageref{firstpage}--\pageref{lastpage}}
\maketitle

\begin{abstract}
  Forthcoming cosmic shear surveys will make precise measurements of the matter
  density field down to very small scales, scales which are dominated by
  baryon feedback. The modelling of baryon feedback is crucial to ensure
  unbiased cosmological parameter constraints; the most efficient approach
  is to use analytic models, but these are limited by how well they can capture the physics of baryon feedback.
  We investigate the fitting and residual errors of various baryon feedback 
  models to a suite of hydrodynamic simulations, and
  propagate these to cosmological parameter constraints for cosmic shear. We
  present an alternative formalism to binary scale-cuts through the
  use of a theoretical error covariance, which is a well-motivated alternative
  using errors in the power spectrum modelling itself. We depart from previous
  works by modelling baryonic feedback errors directly in the matter power 
  spectrum, which is the
  natural basis to do so and thus preserves information in the lensing kernels.
  When including angular multipoles up to $\lmax = 5000$, and assuming \textit{Euclid}-like
  survey properties, we find that even multi-parameter models of baryon feedback can
  introduce significant levels of bias. In contrast, our theoretical error 
  reduces the bias in $\Omegam$ and $\Seight$ to acceptable levels, with only a modest increase in 
  parameter variances. The theoretical error approach bypasses the need to directly
  determine the per-bin $\lmax$ values, as it naturally suppresses the biassing 
  small-scale information. We also present a detailed study of how flexible 
  \HMCode-2020, a widely-used non-linear and baryonic feedback model, is at
  fitting a range of hydrodynamical simulations.
\end{abstract}



\section{Introduction}
\label{sec:Introduction}

Cosmic shear is the coherent distortion in the apparent shapes of galaxies due 
to the matter distribution of the large-scale structure of the universe 
\citep{Bartelmann:1999yn,Bartelmann:2010fz,Kilbinger:2014cea}. These 
distortions are sensitive to the total matter inhomogeneities along the 
line-of-sight, and thus are a powerful probe of the non-luminous dark matter, which
ordinarily cannot be directly observed using telescopes.
Cosmic shear is also sensitive to the detailed physics of baryons, particularly on small
scales, within the Universe. By making accurate measurements, 
along with robust theoretical modelling, cosmic shear is able to provide
detailed knowledge about the physics and geometry of the universe.

Cosmic shear surveys have already placed tight constraints on the fundamental physics
and properties of our universe, especially on the growth of structure parameter
$\Seight$, defined as $\Seight \equiv \sigmaeight \sqrt{\Omegam / 0.3}$.
The results from existing surveys have been well-studied, with data
coming from the Kilo-Degree Survey (KiDS-1000) \citep{Heymans:2020gsg,KiDS:2020suj,Li:2023azi},
the Dark Energy Survey (DES-Y3) \citep{DES:2021wwk,DES:2021bvc,DES:2021vln,DES:2022qpf},
and Hyper Suprime-Cam (HSC-Y3) \citep{Li:2023tui,Dalal:2023olq}.

Over the next decade, an unprecedented amount of high-quality cosmic shear data
will be released. This will come from the recently launched \textit{Euclid} 
space telescope \citep{Euclid:2011zbd,Euclid:2024yrr}, the Legacy Survey of Space and Time
(LSST) at the \textit{Rubin} observatory \citep{LSSTDarkEnergyScience:2012kar}, 
and the \textit{Roman} space telescope \citep{Spergel:2015sza}. Since the
quality and quantity of cosmic shear data that is expected to be produced is
so vast, the accuracy of the theoretical modelling is required to be as equally
precise. 

While the majority of the matter in the universe is dark matter, which only
interacts gravitationally, the baryons in the universe, while appearing to be
lighter in total mass, have an equally large affect on the dynamics of the universe --
particularly on small scales. Baryons are responsible for the heating and
cooling of gas, the creation and demise of stars, and play an important part in
the feedback from Active Galactic Nuclei (AGN), which can have considerable
impact on the matter power spectrum over a wide range of scales \citep{vanDaalen:2011xb}.

The modelling of baryon feedback and its effects on cosmic shear cosmology
has been discussed extensively in the 
literature, with very many different methods and implementations proposed
to mitigate its effects. These include \citet{Schneider:2015wta},
\citet{Giri:2021qin}, \citet{Arico:2020lhq}, \citet{Huang:2018wpy}, \citet{Salcido:2023etz}
and \citet{Mead:2020vgs}, among many others. Cosmic shear is a measurement
in angular-space due to the necessity of using coarse photometric redshift
estimates, and so this makes the measurements of cosmic shear sensitive to
high-$k$ wavenumbers throughout any $\ell$ values in the angular power spectrum.  
It is at high $k$ where feedback effects become relevant.

A popular approach to mitigating baryonic effects in cosmic shear 
analyses is to introduce scale-cuts into the data-vector. Here, physically 
smaller scale elements in the data vector beyond a cut-off, either in
real-space $\theta_{\textrm{min}}$ or in Fourier-space $\ell_{\textrm{max}}$,
are completely discarded even if high signal-to-noise observations have been 
taken. These scale-cuts are often dependent on the redshift of the source 
galaxies, as for further away redshift bins the same physical scale is given 
by larger $\theta$ or smaller $\ell$. One issue that arises with such scale-cuts
is that it by definition is a \textit{hard cut} on the data. For example, 
if it was chosen that $\ell_{\textrm{max}} = 2000$, then $\ell = 2000$ would
be included in an analysis whereas the $\ell = 2001$ mode would be excluded,
even though these modes will be highly coupled and have very similar
theoretical uncertainties.
The use of these binary scale cuts have been discussed heavily in previous
cosmic shear results, most recently in the Dark Energy Survey's Year 3 
results (real-space) \citep{DES:2021rex,DES:2021wwk},
the Kilo-Degree Survey fourth data release (KiDS-1000, in real-space) \citep{Li:2023azi},
and Hyper Suprime-Cam's Year 3 results
(harmonic-space) \citep{Dalal:2023olq}. 

Since the use and location of a hard scale cut could be considered somewhat
unphysical, in the sense that we have two consecutive data points one with
little error and one with infinite error, we investigate and present results for
an alternative: the use of a theoretical error covariance which acts as a soft
scale cut that is informed by our inability to correctly model baryonic feedback
down to arbitrarily small scales. This has been explored in the literature 
previously, with the theory being originally presented in 
\citet{Baldauf:2016sjb}, expanded upon in \citet{Sprenger:2018tdb},
and applied to mock cosmic shear analyses in
\citet{Moreira:2021imm}, among others~\citep{Pellejero-Ibanez:2022efv}.
Baryonic feedback effects directly impact
the matter power spectrum, which is then integrated over to produce effects in
the lensing angular power spectrum. Thus, when we model our theoretical
uncertainties coming from baryonic feedback, it is most natural to do so in
the matter power spectrum and propagate these uncertainties to the angular
power spectrum. \citet{Moreira:2021imm} investigated theoretical uncertainties
with respect to the \textit{angular} power spectrum, and so we are revisiting this
formalism but extending it to the underlying matter power spectrum. Since
baryonic feedback contaminates only the matter power spectrum, not the lensing
kernels associated with cosmic shear, isolating the errors associated with
the matter power spectrum then propagating these to the angular power spectrum
is a sensible alternative approach. The theoretical uncertainties approach is
similar to analytic marginalisation of small-scale physics, which also
results in an additional term to the covariance matrix~\citep{Kitching:2010ab}.

This paper is structured as follows: in Section~\ref{sec:modelling_baryons} we
outline the need for analytic baryon feedback models and the use of hydrodynamical
simulations, Section~\ref{sec:Methodology} discusses our methodology for
constructing our theoretical error covariance, in Section~\ref{sec:results}
we present our results for cosmological parameter constraints using a selection
of baryon feedback models, and Section~\ref{sec:diss_and_conc} summarises
our findings.

\section{Modelling baryonic feedback in the matter power spectrum}
\label{sec:modelling_baryons}

\begin{figure*}
  \includegraphics[width=2\columnwidth]{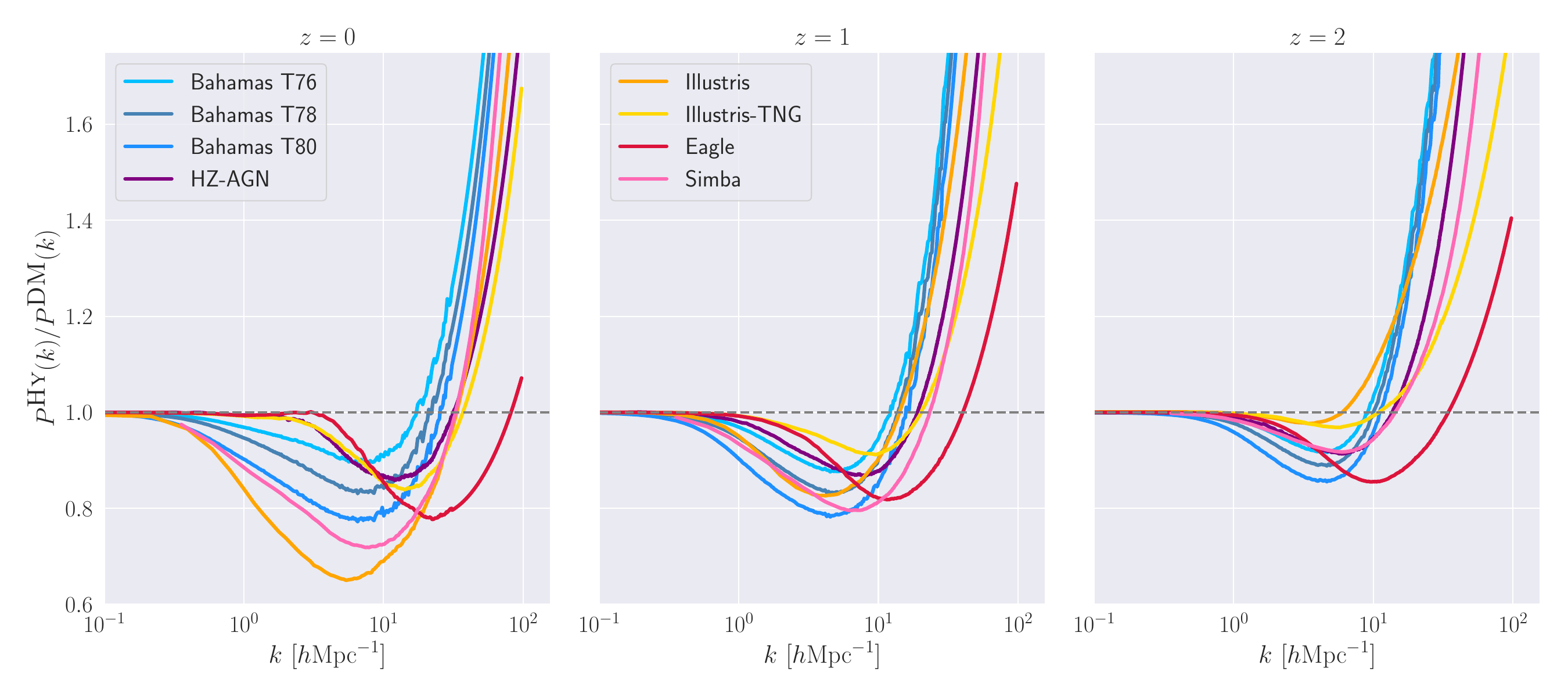}
  \vspace*{-0.25cm}
  \caption{Ratios of the matter power spectrum with baryonic feedback in to
    their dark-matter-only counterparts at redshifts 0, 1, and 2, for various
    hydrodynamical simulations, as given in Table~\ref{tbl:hydrosims}.
    Here, `\Bahamas~T$XY$' corresponds to the same underlying \Bahamas
    hydro-sim, just run at different $\Delta T_{\textrm{heat}}$ values, which
    corresponds to the amplitude of AGN feedback within their simulation~\citep{McCarthy:2016mry}.
    Note that \Simba was run on a smaller box-size with respect to the
    other simulations (as shown in Table~\ref{tbl:hydrosims}), and thus
    we do not have access to data below a certain $k$ for \Simba, which is much
    larger than the smallest $k$ modes available for the other hydro-sims.
    }
  \label{fig:My_Pk_ratio}
\end{figure*}

Accurate evaluations of the non-linear dark-matter-only matter power spectrum is
of extreme importance to any cosmic shear analysis \citep{Schneider:2015yka}. 
There are numerous ways to accurately evaluate this. For example, many models
descend from the original halo model presented in the early 2000s
\citep{Seljak:2000gq,Peacock:2000MNRAS3181144P,Cooray:2002dia,Smith:2002dz}:
the \Halofit model presented in ~\cite{Takahashi:2012em}, and \HMCode models
that originate with \HMCode-2015~\citep{Mead:2015yca}, then \HMCode-2016~\citep{Mead:2016zqy},
and most recently \HMCode-2020~\citep{Mead:2020vgs}. All these models serve to
provide accurate predictions for the non-linear dark-matter-only matter power
spectrum as a function of cosmological parameters.

The most accurate way of modelling baryonic feedback physics is through
full hydrodynamical cosmological simulations (hydro-sims)~\citep{vanDaalen:2011xb,Vogelsberger:2019ynw}.
There are many different hydro-sims within the literature, which mainly differ
in their implementations of sub-grid physics. There are suggestions that these
sub-grid physics differences are connected to the gas fraction in galaxy
groups and clusters~\citep{Salcido:2024qrt}. These simulations are extremely
computationally expensive due to the need for
large volumes, to reduce cosmic variance and capture clustering on cosmological 
scales, and the need to model small-scale behaviour, which directly impacts the
larger-scale clustering. Thus, these simulations need sufficient resolution else
the baryonic feedback physics which we wish to capture will simply be washed out
on larger scales and also sufficient volume. This presents a challenging
dynamic range problem, hence the very many different simulations that 
implement these range of physics.

To isolate the effect of baryonic physics on the matter power spectrum when
compared to the dark-matter-only non-linear power spectrum, we use the baryon
response function, $R(k, z)$, defined as
\begin{align}
  R(k, z) \equiv \frac{\Phykz}{\Pdmkz},
  \label{eqn:baryonic_pk_ratio}
\end{align}
which isolates the effects of baryon physics from the general non-linear
spectrum, and is especially useful for reducing the effects of cosmic variance
on the power spectrum when using hydro-simulations. This ratio is approximately
insensitive to the cosmological parameters that $\Phy$ and $\Pdm$ were
generated using, provided that they are the same cosmology~\citep{vanDaalen:2019pst,Elbers:2024dad}.
This makes the baryon response function especially useful when trying to compare predictions between different
hydro-sims run at different cosmologies. Since every hydro-sim has a different 
implementation of their astrophysical processes, the range of curves for
$R(k, z)$ can be extremely broad. Figure~\ref{fig:My_Pk_ratio} plots this
ratio for a verity of different hydro-sims for three redshifts. This shows the
general behaviour of $R(k, z)$ that has been well explored previously~\citep{vanDaalen:2011xb}:
there is a dip in expected power on non-linear scales 
($1 \, h\Mpc^{-1} \lesssim  k \lesssim 10 \, h\Mpc^{-1}$ at $z=0$ )
due to baryon feedback processes, such as AGN, jets, and supernovae,
expelling matter reducing clustering, while the dramatic
increase on highly non-linear scales ($ k \gtrsim 10 \, h\Mpc^{-1}$) is due to
additional clustering from the ability of baryons to undergo radiative cooling,
increasing small-scale clustering, and from star formation within the simulations~\citep{Chisari:2019tus}.
Also shown in Figure~\ref{fig:My_Pk_ratio} is the level of scatter in the prediction of $R(k, z)$
for different hydro-sims, with the depth and location of the suppression and
location of the upturn highly simulation dependant with very little common 
consensus between these simulations. \cite{vanDaalen:2019pst} has shown that
much of this scatter in the predictions for the suppression in the simulations can be
explained by how the amplitude of the suppression is strongly correlated with
the mean baryon fraction in haloes of mass~$\sim \!\! 10^{14} M_{\odot}$~\citep{Salcido:2024qrt}.

Due to the large scatter in the predictions of the hydro-sims, we take an 
agnostic approach to their results: we assume that each of these simulations
are equally trustworthy, and thus any of these predictions for their $R(k, z)$
must be equally reliable. We must also assume that the true
description of our Universe lies somewhere in this ensemble in order to correctly
match observational data.
Hence, any baryon physics model must be able to recreate all of these curves
in order for us to sufficiently trust that it could potentially capture the real
physics of our Universe. This level of scatter in the hydro-sims has lead to the development of many
baryonic physics models, especially those which aim to match the simulations
through neural-network emulator methods, such as the \Bacco emulator presented
in~\cite{Arico:2020lhq}. \Bacco claims accuracy 1-2$\%$ for scales
$1 \, h\Mpc^{-1} <  k < 5 \, h\Mpc^{-1}$ and redshifts $0 < z < 1.5$, and again
finds that the most important parameter that controls baryon feedback physics
is the gas fraction per halo mass. Furthermore, \BCEmu~\citep{Giri:2021qin}
implements the `baryonification model' prescription~\citep{Schneider:2018pfw}
which claims percent-level accuracy for scales below $k \sim 10 \, h\Mpc^{-1}$
at redshifts $z < 2$.

\begin{figure}
  \includegraphics[width=\columnwidth]{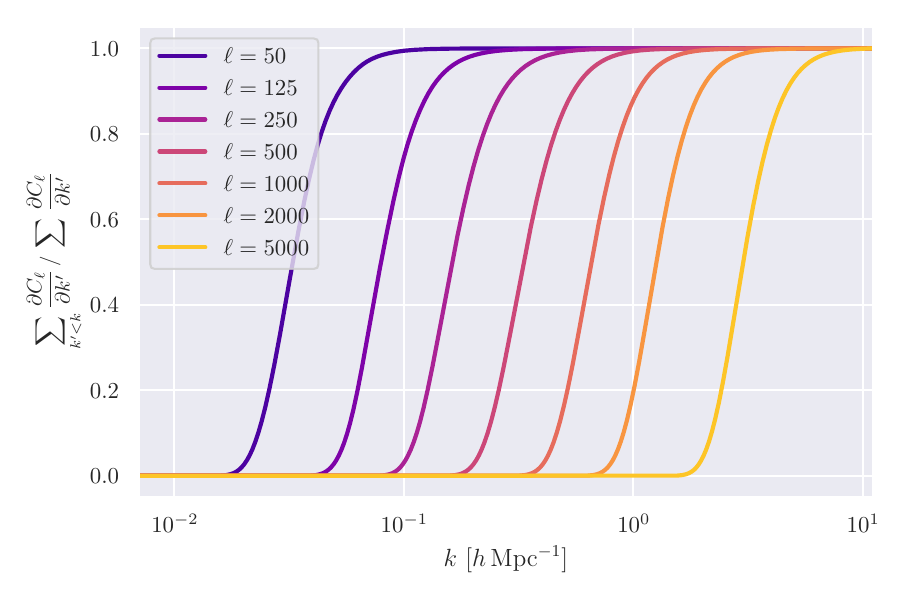}
  \vspace*{-0.25cm}
  \caption{Cumulative sum of the derivate of the cosmic shear power angular
    power spectrum coefficients $\Cl$, at certain $\ell$ modes, with respect to
    the wavenumber $k$ in the Limber integral (Equation~\ref{eqn:cosmic_shear_powspec}).
    The sum of the derivatives for each multipole is normalised to unity for
    easy comparison between different modes.
    This includes the contributions from the lensing kernels and the matter power spectrum. Derivatives were
    taken for a Gaussian source redshift distribution located at $\bar{z} = 0.33$
    and width $\sigma_{z} = 0.15$. }
  \label{fig:Cl_deriv_cumsum}
\end{figure}

We are motivated to find correct descriptions of baryon feedback physics due to
their large impact on results obtained from cosmic shear surveys. Future
Stage-IV cosmic shear surveys hope to probe extremely small scales, up to a
maximum multipole of $\lmax \sim 5000$ \citep{Euclid:2011zbd}. These small
angular scales probe the highly non-linear regime in the matter power spectrum,
as shown in Figure~\ref{fig:Cl_deriv_cumsum}. Here, we plot the cumulative sum
of the derivate of the angular power spectrum coefficients $\Cl$ at certain
$\ell$ values as a function of $k$. This shows the relative contribution for
each $k$ mode to each $\Cl$ value. Hence, we find that if we hope to probe
angular scales up to $\ell = 5000$, then we need to have a strong understanding of
the matter power spectrum up to scales $k \sim 10 \, h\Mpc^{-1}$ -- including
the details of baryonic physics. If these scales are not correctly modelled,
then this could induce catastrophic biases in a cosmic shear analysis~\citep{Semboloni:2011fe,Huang:2018wpy}, 
particularly for the constraints on dark energy~\citep{Copeland:2017hzu}.
The alternative is to reduce the maximum multipole $\lmax$ in our analyses,
though this discards potentially useful cosmological information.

This motivates us to test the accuracy of baryonic feedback models up to these
small scales, and assess how different treatments of baryonic physics impact
constraints obtained from forthcoming cosmic shear surveys.

\subsection{Theoretical error formalism}
\label{sec:theory_error_formalism}

Since we do not have models of baryonic physics that completely match
our complete set of hydrodynamical simulations, there will always be some
residual error for any given model of baryon feedback physics as a function of
scale and redshift. Rather than treating our models as a perfectly known 
quantity and limiting our scales of interest, we can directly incorporate these
known errors through the use of the `theoretical uncertainty' formalism 
first presented in~\cite{Baldauf:2016sjb}. We now have a theoretical error
data-vector $\mathbfit{e}$, which quantifies the difference between the true, underlying
physical model and what our approximative methods calculate, which is bound
by an envelope $\mathbfit{E}$. This theoretical error has its associated
covariance matrix, $\mathbfss{C}^{\textsc{e}}$, which can be written as a
function of the amplitude of the error, $\mathbfit{E}$, and a correlation matrix
$\boldsymbol{\rho}$ as
\begin{align}
  \mathbfss{C}^{\textsc{e}} = \mathbfit{E} \, \boldsymbol{\rho} \, \mathbfit{E}.
\end{align}
In the Gaussian approximation, we can simply model the inclusion of the 
theoretical error on the likelihood as simply the addition of our theoretical
error covariance, $\mathbfss{C}^{\textsc{e}}$, to our existing data covariance
matrix (which includes the contributions from cosmic variance),
$\mathbfss{C}^{\textsc{d}}$, to give
\begin{align}
  \mathbfss{C}^{\textsc{tot}} = \mathbfss{C}^{\textsc{d}} + \mathbfss{C}^{\textsc{e}}.
  \label{eqn:cov_addition}
\end{align}
The theoretical error formalism can be, in principle, applied to any effect
that changes summary statistics that is not yet perfectly modelled.
This approach of marginalising over the errors in the model was first applied
to baryonic physics effects for cosmic shear surveys in \cite{Moreira:2021imm},
where they modelled the errors and correlations directly in $\ell$-space.
They constructed several versions of the error envelope function $E(\ell)$,
all featuring the similar term of the error ratio of best-fit $\Cl$ values
to the values generated from hydro-sims. They also assumed a Gaussian-like
function in $(\ell - \ell')^2$ for their correlation matrix, with a characteristic
correlation length~$L$ determining the widths of the correlation matrix. 

Since baryonic effects directly change the matter power spectrum, which is then
propagated to cosmic shear summary statistics, we are motivated to investigate
the theoretical errors in $k$-space for the matter power spectrum and then
propagate those to $\ell$-space. There is no theoretical error associated with
the lensing kernels, so by isolating the theoretical uncertainties to the matter
power spectrum only, we might hope to preserve information from geometry this way.
The task now becomes one to find an appropriate
theoretical error envelope function $\mathbfit{E}$ and correlation 
matrix $\boldsymbol{\rho}$ in $k$-space.

\section{Methodology}
\label{sec:Methodology}

\subsection{Modelling forthcoming cosmic shear surveys}

We use the standard prescription for the cosmic shear power 
spectrum~\citep{Bartelmann:1999yn,Bartelmann:2010fz,Kilbinger:2014cea} where
the power spectrum is, for tomographic redshift bins labelled by $a$ and $b$, is
given as
\begin{align}
  \Cl^{ab} = \frac{9}{4} \frac{\Omegam^2 H_0^4}{c^4} \int_{0}^{\chi_{\textrm{max}}}
  \!\! 
  \d \chi \! \frac{g_{a}(\chi) \, g_{b}(\chi)}{a^2(\chi)} \, P \! \left(k=\frac{\ell}{\chi}, \, z=z(\chi) \right),
  \label{eqn:cosmic_shear_powspec}
\end{align}
where $a(\chi)$ is the scale factor, $P$ is the non-linear matter power spectrum,
and $g(\chi)$ is the lensing kernel given as
\begin{equation}
  g_{a}(\chi) = \int_{\chi}^{\chi_{h}} \!\! \d \chi' \, n_{a}(\chi') \frac{f_{K}(\chi' - \chi)}{f_{K}(\chi')},
\end{equation}
where $n(\chi)$ is the probability density of source galaxies as a function of
comoving distance. To evaluate the non-linear matter power spectrum with baryonic
feedback, we used \HMCode-2020 with $\Tagn = 7.8$.

Since every galaxy has an intrinsic ellipticity, this introduces a shape
noise term into the power spectrum with a flat value $N_{\ell}$ given as\
\begin{align}
  N_{\ell}^{ab} = \frac{\sigma_{\epsilon}^2}{\bar{n}} \, \delta^{ab}, 
\end{align}
where $\sigma_{\epsilon}$ is the standard deviation of the intrinsic galaxy 
ellipticity dispersion per component, $\bar{n}$ is the expected number of
observed galaxies per steradian, and $\delta^{ab}$ is the Kronecker-$\delta$
symbol. We assume \textit{Euclid}-like values where it is expected that 
30 galaxies per square arcminute will be observed, but we divide these into five
photometric redshift bins, giving 
$\bar{n} = 6\,\mathrm{gals / arcmin}^{2}$~\citep{Euclid:2011zbd}. 
We take $\sigma_{\epsilon} = 0.21$. 

We use five Gaussian redshift bins with means located at 
$\bar{z} = \{0.33, 0.66, 1.0, 1.33, 1.66\}$ all with standard deviation of
$\sigma_z = 0.15$. This gives us 15 unique power spectrum combinations and
120 unique covariance matrix blocks.

We model the Gaussian covariance matrix as the four-point function~\citep{Euclid:2019clj},
given by
\begin{align}
  \textrm{Cov}[\Cl^{ab}, \Clp^{cd}] = 
  \frac{\Cl^{ac} \Cl^{bd} + \Cl^{ad} \Cl^{bc}}{(2 \ell + 1) \, \fsky} \delta_{\ell \ell'},
  \label{eqn:Gaussian_Cl_cov}
\end{align} 
where $\fsky$ is the fraction of sky observed by the cosmic shear survey. 
We take a \textit{Euclid}-like value of $\fsky = 0.35$.

To reduce the dimensionality of our power spectrum and covariance matrices, we
bin the power spectrum. We use linear binning of five bins up to $\ell = 100$,
and then use logarithmic binning with twenty bins from $\ell = 100$ to 
$\ell = 5\,000$. We use an $\ell$-mode weight of $\ell(\ell+1)$ when binning.

\subsection[Constructing our k-space theoretical covariance]{\boldmath Constructing our $k$-space theoretical covariance}

As outlined in Section~\ref{sec:theory_error_formalism}, the theoretical error
formalism requires us to quantify an error envelope function $\mathbfit{E}$ and
a correlation matrix $\boldsymbol{\rho}$, and since our uncertainties from
baryonic physics are best specified in $k$-space, we will first construct our
$k$-space covariance matrix. We will first discuss the construction of our
envelope function $\mathbfit{E}$.

There is no absolute choice for the functional form of $\mathbfit{E}$.
\cite{Moreira:2021imm} investigated many choices for the form of $\mathbfit{E}$,
with the `mirror' envelope being a good fit to the data. Here, we follow their
method and take the maximum deviation for each $k$-mode and redshift across a
suite of hydrodynamical simulations to the best-fitting values of a baryon 
feedback model (see Section~\ref{sec:fitting_hmcode_to_hydrosims}).
This gives the maximum relative difference in the baryonic
response function (Equation~\ref{eqn:baryonic_pk_ratio}) as
\begin{align}
  \Delta R(k, z) = \max_{k, \, z} \left\lvert \frac{R^{\textsc{Hydro}}(k, z)}{R^{\textsc{Model}}(k, z)} - 1 \right\rvert.
  \label{eqn:max_deviation}
\end{align} 
We then turn this into an amplitude by multiplying by a fiducial 
dark-matter-only power spectrum,
\begin{align}
  E(k, z) = P^{\textsc{dm}}(k, z) \, \Delta R(k, z).
\end{align}
where $\Pdmkz$ was evaluated using the \HMCode-2020 DM-only model.

With our amplitude function specified, we now wish to construct our $k$-space
correlation matrix. Here, we are considering the correlations between two
pairs of points $(k_1, z_1)$ and $(k_2, z_2)$. We follow previous works that
have employed the theoretical uncertainties approach (for example \cite{Baldauf:2016sjb,Sprenger:2018tdb,Chudaykin:2019ock,Moreira:2021imm})
by choosing a factorisable Gaussian correlation
matrix of the form
\begin{align}
  \rho[(k_1, z_1), \, (k_2, z_2)] = 
  \exp \left[ -\frac{\log(k_1 / k_2)^2}{\sigma_{\log k}^2} \right] 
  \exp \left[ -\frac{(z_1 - z_2)^2}{\sigma_{z}^2} \right],
\end{align}
where $\sigma_{\log k}$ and $\sigma_z$ are the characteristic correlation scales in
$\log k$- and redshift-space, respectively. In our fiducial analyses, we used
values of $\sigma_{\log k} = 0.25$ and $\sigma_z = 0.25$. These values were motivated
from Figure~\ref{fig:My_Pk_ratio} where we see that the baryonic response
function $R(k, z)$ is correlated on scales of approximately a quarter of a
decade in log-$k$ space with approximately the same coupling in redshift-space.
We discuss the effects of changing $\sigma_{\log k}$ and $\sigma_z$ in Section~\ref{sec:sigma_kz}.

We are motivated to use factorisable
Gaussians since we know that baryon feedback has a relatively local effect on
the matter power spectrum, that is neighbouring wavenumbers and redshifts 
behave similarly and thus should have highly correlated covariances, whereas
vastly different wavenumbers and redshifts have very different evolutionary 
physics, and thus should be less correlated. Furthermore, we are motivated to 
use the logarithmic differences in $k$-space since our wavenumbers span many
orders of magnitude (see Figure~\ref{fig:My_Pk_ratio}), and thus an ordinary
difference might not properly reflect this. 

Combining our envelope function and correlation matrix, we find that our
complete $k$- and redshift-space covariance matrix is given by 
\begin{multline}
  \textrm{Cov}[(k_1, z_1), (k_2, z_2)] = \\
  P^{\textsc{dm}}(k_1, z_1) \, \Delta R(k_1, z_1) \,\,  P^{\textsc{dm}}(k_2, z_2) \, \Delta R(k_2, z_2) \, \times \\
  \exp \left[ -\frac{\log(k_1 / k_2)^2}{\sigma_{\log k}^2} \right] \times \exp \left[ -\frac{(z_1 - z_2)^2}{\sigma_{z}^2} \right].
  \label{eqn:k_space_cov}
\end{multline}

We note that by working directly with errors in the matter power spectrum, which is the
underlying quantity that we have limited knowledge in modelling (not the 
angular power spectrum values), this should capture the full phenomenology of 
the errors in baryonic feedback effects on the matter power spectrum. This 
contrasts with the work of \cite{Moreira:2021imm} which also aims to mitigate
baryonic physics effects through the use of a theoretical uncertainty, though
they quantify their theoretical uncertainty through the differences in
the angular power spectrum values.

\subsection[Propagating covariances to ell-space]{\boldmath Propagating covariances to $\ell$-space}

With our $k$- and redshift-space covariance matrix specified, we can now 
propagate this into an $\ell$-space covariance matrix, which can then be added
to the data covariance to give our overall covariance matrix.

Equation~\ref{eqn:cosmic_shear_powspec} gave the cosmic shear power spectrum
as the Limber integral of the matter power spectrum weighted by the geometrical
factors.  Hence, if we now want to propagate our uncertainties in the $P(k)$ into an
additional covariance matrix for the $\Cl$ values, we can integrate this again, 
which yields
\begin{multline}
  \textrm{Cov}[C_{\ell_1}^{ab}, \, C_{\ell_2}^{cd}] =
  \left[ \frac{9}{4} \frac{\Omega_m^2 H_0^4}{c^4} \right]^2 \times
  \int_{0}^{\chi_{\textrm{max}}} \!\! \int_{0}^{\chi_{\textrm{max}}}  \!\! \d \chi_1 \d \chi_2 \\
  \frac{g_a(\chi_1) \, g_b(\chi_1)}{a^2(\chi_1)} \, \frac{g_c(\chi_2) \, g_d(\chi_2)}{a^2(\chi_2)} \,\,
  \textrm{Cov}[(k_1, z_1), \, (k_2, z_2)],
  \label{eqn:cov_dbl_integral}
\end{multline}
where $k_1 = \frac{\ell_1}{\chi_1}$ and $k_2 = \frac{\ell_2}{\chi_2}$. 
We are motivated to use Limber's approximation here to simplify the double
integral since the low $\ell$ region in which the approximation is imprecise 
will have a negligible contribution to the total covariance.

\subsection{Numerical evaluation of the matter power spectrum with baryon 
feedback}

Our construction of the $(k, z)$-space covariance matrix can be applied to any
model of baryon feedback, with models ranging from purely analytical methods
using the Zel'dovich approximation~\citep{Mohammed:2014lja}, to semi-analytic
models (e.g. \HMCode, \citealt{Mead:2020vgs}), to purely numerical emulation (e.g.
\Bacco, \citealt{Arico:2020lhq} and \BCEmu, \citealt{Schneider:2018pfw}).

We have chosen \HMCode-2020 as our model of choice for the matter power spectrum
and baryonic feedback response since they claim that their dark-matter-only spectrum
has RMS errors of less than $5$ per-cent, and that their 
baryonic feedback response is accurate to within the $1$ per-cent level for redshifts 
$z < 1$ and scales ($k < 20 \, h\textrm{Mpc}^{-1}$), over a range of cosmologies~\citep{Mead:2020vgs}.
Thus, \HMCode-2020 is a natural choice for computing the
matter power spectrum for use in cosmic shear analyses, which is confirmed
by its use in recent major cosmic shear results such as the joint \KiDS--\DES
analysis \citep{Kilo-DegreeSurvey:2023gfr}.

\HMCode-2020 comes in two distinct flavours for modelling baryon physics:
\begin{itemize}
  \item A full six-parameter description with free-parameters of the halo 
    concentration, $B$, the stellar mass fraction in haloes, $f_\star$, and the
    halo mass break scale, $\Mb$, along their redshift evolution counterparts.
    The redshift evolution for each physical parameter $X$ is given by 
    the fixed form of $X(z) = X_0 \times 10^{z \, X_z}$. 
  \item A one-parameter version where the feedback temperature $\Tagn$
    encapsulates the combined baryon physics of the full model. The 
    one-parameter model was constructed by fitting a linear relationship to 
    the six-parameter version using the \Bahamas simulations. Since the 
    one-parameter model was specifically constructed to the \Bahamas
    simulations, there is no guarantee of its accuracy to other hydrodynamical
    simulators or even the baryon physics of our own Universe.
  \item We also test a three-parameter version, which is analogous to the 
    six-parameter model but with fixed redshift evolution. This is a useful
    test as there arises significant degeneracies between the amplitude and
    redshift evolution of each parameter, and thus by fixing its evolution we
    can test if this extra degree of freedom is necessary.  
\end{itemize} 

While a more general description of baryon feedback physics should have more freedom
to better match different physical models, it comes at the cost of additional
nuisance parameters which needs to be included and marginalised over in any
analysis. This could slow down the convergence of analysis pipelines, and in
the case of very many parameter models, unnecessarily decreasing constraints
on  the core cosmological parameters due to the need to excessively marginalise
over these baryonic nuisance parameters. Hence, while we are focusing on the
inclusion of theoretical uncertainties into our analysis pipeline, we also
look at how going from a one- to three- to six-parameter baryonic feedback
model changes our results.

\subsection{Chosen hydrodynamical simulations}

We make use of the `power spectrum library' presented 
in~\citet{vanDaalen:2019pst}\footnote{\url{https://powerlib.strw.leidenuniv.nl}}
and from the \Camels Project~\citep{CAMELS:2023wqd}\footnote{\url{https://camels.readthedocs.io/}}
to build up a suite of six hydrodynamical simulations (with three flavours of
the Bahamas simulations) as shown in Table~\ref{tbl:hydrosims}. This gives us a
wide range of baryon feedback models against which we can benchmark our analytic 
models to.

\begin{table*}
  \caption{The six hydrodynamical simulations used in this work to benchmark
    and compare models of baryon feedback physics to.
    }
  
  \label{tbl:hydrosims}
  
  \begin{tabular}{lrrrrr}
    \hline

    Hydro-sim & Box-size (Mpc) & Number of particles & Baryonic mass resolution $[M_{\odot}]$ & Cosmology & Reference\\
    
    \hline
    Bahamas           & $400 / h$ & $2 \times 1024^{3}$ & $7.66 \times 10^{8} \, h^{-1}$ &  \textit{WMAP} 9 & \cite{McCarthy:2016mry} \\

    Horizon-AGN       & $100 / h$ & $1024^{3}$ & $8.3\times 10^{7} $ & \textit{WMAP} 7 & \cite{Chisari:2018prw} \\

    Illustris         & 106.5     & $1820^{3}$ & $1.6 \times 10^{6}$ & \textit{WMAP} 7 & \cite{Nelson:2015dga} \\

    Illustris-TNG 100 & $75 / h$  & $2 \times 1820^{3}$ & $9.44 \times 10^{5} \, h^{-1}$ &\textit{Planck} 2015 & \cite{Springel:2017tpz}\\

    Eagle             & 100       & $1504^{3}$ & $1.81 \times 10^{6}$ &  \textit{Planck} 2013 & \cite{Schaye:2014tpa} \\
    
    Simba             & $25 / h$  & $1024^{3}$ & $ 2.85 \times 10^{5}$ & \textit{Planck} 2013 & \cite{Dave:2019yyq} \\
    
    \hline
  \end{tabular}
\end{table*}

\subsection{Fitting HMCode to hydrodynamical simulations}
\label{sec:fitting_hmcode_to_hydrosims}

Armed with our three baryon feedback models and a suite of hydrodynamical 
simulations, we can now go about performing a best-fit analysis of our models
to the hydro-sims in order to quantify each model's maximum deviation as a 
function of wavenumber and redshift (Equation~\ref{eqn:max_deviation}). To do 
so, we performed a maximum likelihood fit where we simultaneously fitted 
across redshifts $z \leq 2$ with flat weighting in redshifts,
and wavenumbers over the range 
$0.01 \, h\textrm{Mpc}^{-1} \leq k \leq 20 \, h\textrm{Mpc}^{-1}$
with a $k^2$ weighting on each wavenumber to roughly approximate the cosmic 
variance contribution. We are free to choose an arbitrary $k$-mode weighting 
since our envelope function is arbitrary, and so there is no correct choice
for either a $k$- or $z$-mode weight here. We experimented with flat, and a
pure cosmic variance weight of $k^3$, and settled on our $k^2$ weighting
as somewhere in between. Note that, since we use equal spacing in $\log(k)$,
we find $\Delta \log k = \left[\Delta k\right] / k$ when summing over $k$.
This gives our loss function for each model and hydro-sim as
\begin{align}
  \mathcal{L} = \sum_{k, \, z} 
  k^2 \left(R^{\textsc{Hydro}}(k, z) - R^{\textsc{HMcode}}(k, z) \right) ^ 2.
  \label{eqn:chi_sq}
\end{align}
The \HMCode power spectra were generated at the same cosmology at which the
hydro-sims were run at, with the minimisation routine just varying the
astrophysical baryonic parameters. The \texttt{Minuit} optimiser, a robust
optimiser as part of the \Cosmosis\footnote{\url{https://cosmosis.readthedocs.io/}}
analysis framework~\citep{Zuntz:2014csq}, was used to maximise the fit
for the baryonic parameters. Since we there is freedom in the form of the 
weight function for each $(k, z)$-mode, different choices for the analytic form 
of $\mathcal{L}$ effectively adds little to that freedom. 

\subsection[Fitting the Cl values]{\boldmath Fitting the $\Cl$ values}

To obtain estimates for the biases in cosmological parameters due to baryon
feedback, many MCMC analyses were run. We used a custom \Cosmosis pipeline
using the \texttt{PolyChord} nested sampler~\citep{Handley:2015fda,Handley:2015vkr}
with parameters \texttt{LivePoints = 1000}, \texttt{num\_repeats = 60}, 
\texttt{boost\_posteriors = 10}, and \texttt{tolerance = 0.001} for all analyses.
We sampled only over $\Omegac$ and $\As$ for our cosmological parameters, giving
results in terms of $\Omegam$, $\sigmaeight$, and $\Seight$, since cosmic shear
surveys are most sensitive to these cosmological parameters. Thus,
any induced bias will be greatest in these parameters.
In addition, we also
sample over our one, three, and six baryonic feedback parameters, depending
on the model, with wide priors, as shown in Table~\ref{tbl:baryon_priors}, on these nuisance parameters.

\begin{table}
  \caption{Uniform priors used for the \HMCode baryonic feedback models.
  $\Tagn$ is used only in the one-parameter models, with the three-parameter
  model sampling over just the individual amplitude parameters ($X_0$), and the
  six-parameter model including the redshift dependence too ($X_z$).}
  
  \label{tbl:baryon_priors}
  
  \centering

  \begin{tabular}{lr}
    \hline

    Parameter & Uniform prior \\
    
    \hline
    
    $\Tagn$ & $[4.0, \, 12.0]$ \\

    $B_0$ & $[0.25, \, 7.0]$ \\
    $B_z$ & $[-0.5, \, 0.5]$ \\

    $f_{\star, 0}$ & $[0.0, \, 5.0]$ \\
    $f_{\star, z}$ & $[-5.0, \,  5.0]$ \\
    
    $M_{\textrm{b}, 0}$ & $[5.0, \, 20.0]$ \\
    $M_{\textrm{b}, z}$ & $[-2.5, \, 2.5]$ \\
    
    \hline
  \end{tabular}
\end{table}

We ran analyses with just the ordinary Gaussian covariance matrix for our cosmic
shear fields (Equation~\ref{eqn:Gaussian_Cl_cov}), and with the additional
theoretical error covariances (Equation~\ref{eqn:cov_dbl_integral}) added to
the Gaussian covariance (Equation~\ref{eqn:cov_addition}).

A Gaussian likelihood was used, which has been shown to be 
a good approximation to the full Wishart distribution on the 
cut-sky~\citep{Upham:2020klf,Hall:2022das} under the
assumption of Gaussian-only terms in the covariance matrix, which we are 
employing.

We generate the input cosmic shear $\Cl$ values for each hydro-sim by taking
each hydro-sim's baryon response function $R$ and multiplying it by \HMCode's
dark-matter only matter power spectrum at the hydro-sim's cosmology, giving
\begin{align}
  \Phykz = R(k, z) \, \times \, P^{\textsc{HMcode-DM}}(k, z).
  \label{eqn:baryon_response_func}
\end{align}
Thus, we are isolating the effects of baryon feedback in the $\Cl$ values, and
not studying other effects, for example the details of the non-linear matter
power spectrum.

\section{Results}
\label{sec:results}

\subsection{Results of fitting HMCode to hydro-sims}

\begin{figure}
  \includegraphics[width=\columnwidth]{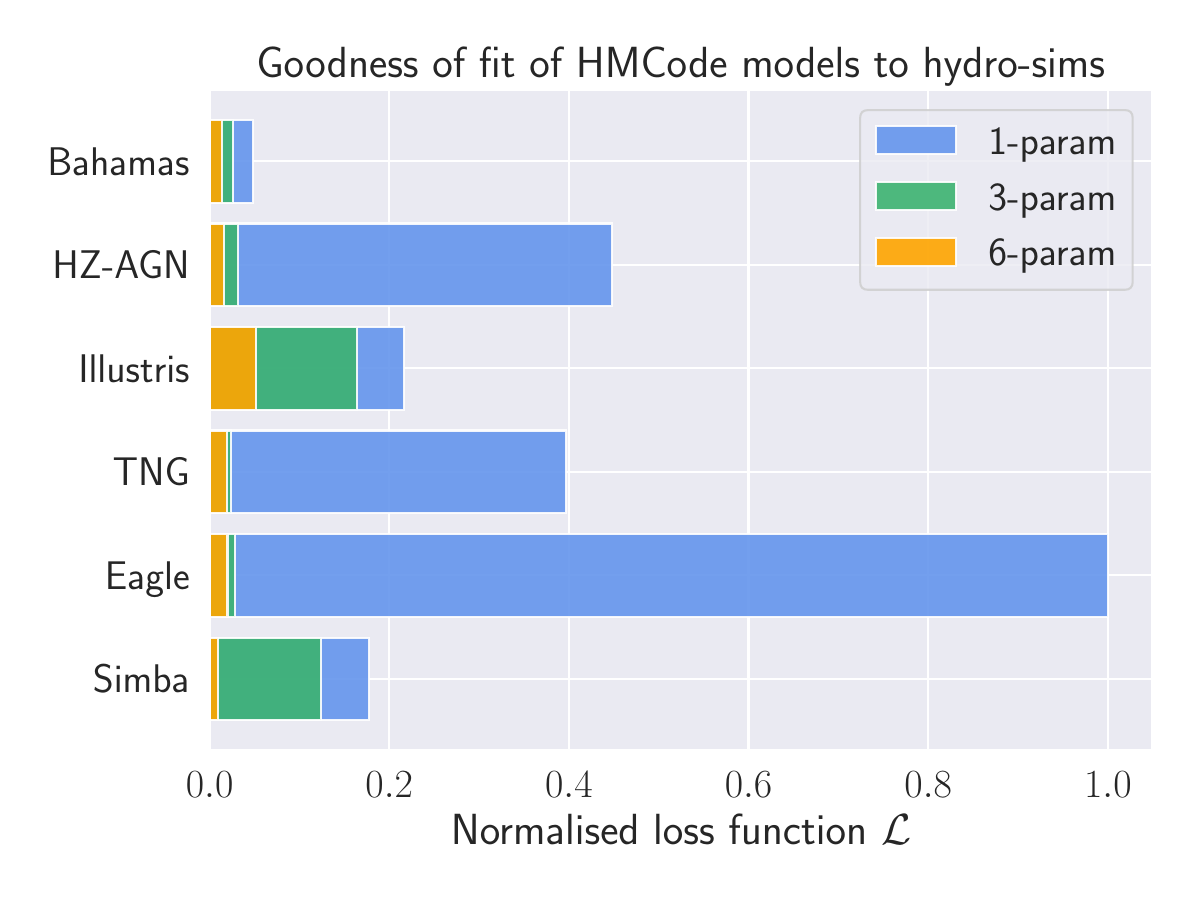}
  \vspace*{-0.5cm}
  \caption{Goodness of fit statistics for the \HMCode 1-, 3-, and 6-parameter
    models for our suite of hydro-simulations. We plot the reduced loss function
    $\mathcal{L}$ values (Equation~\ref{eqn:chi_sq} divided by the total number of data-points
    across our redshift and wavenumber ranges) since each hydro-sim has a
    different number of redshift bins and wavenumbers that the power spectra
    were evaluated at, normalised to the one-parameter model for the Eagle hydro-sim.  
    Here, we see the inflexibility of the 1-parameter model results in
    significant deviations across all simulations (except \Bahamas which it was
    constructed to fit well), which indicates a poor fit to the data. Extending
    the model to 3- and then 6-parameters further increases the goodness
    of fit to simulations, since we open up addition degrees of freedom within
    the model. We note that the feedback within \Illustris is quite extreme, and
    thus produces a degraded fit even with the six-parameter model. 
    }
  \label{fig:HMcode_Pk_bestfit}
\end{figure}

The first task was to fit our three \HMCode baryon physics models to the suite
of hydro-sims. Figure~\ref{fig:HMcode_Pk_bestfit} shows the goodness of fit
statistic for each hydro-sim for our 1-, 3-, and 6-parameter model. We see that 
the 1-parameter model generally provides a poor fit to our hydro-sims due to its
inability to match the general baryon feedback models. When we extend the
generality to 3- and 6-parameters, the goodness of fit increases which matches
our intuition that a more general model with higher degrees of freedom should
do a better job at matching arbitrary data sets. 
It should be noted that
the \HMCode 1-parameter model was specifically designed to closely match the
\Bahamas simulations, and so it is no surprise that it has a better fit to 
\Bahamas than other simulations. 

\begin{figure*}
  \includegraphics[width=2\columnwidth]{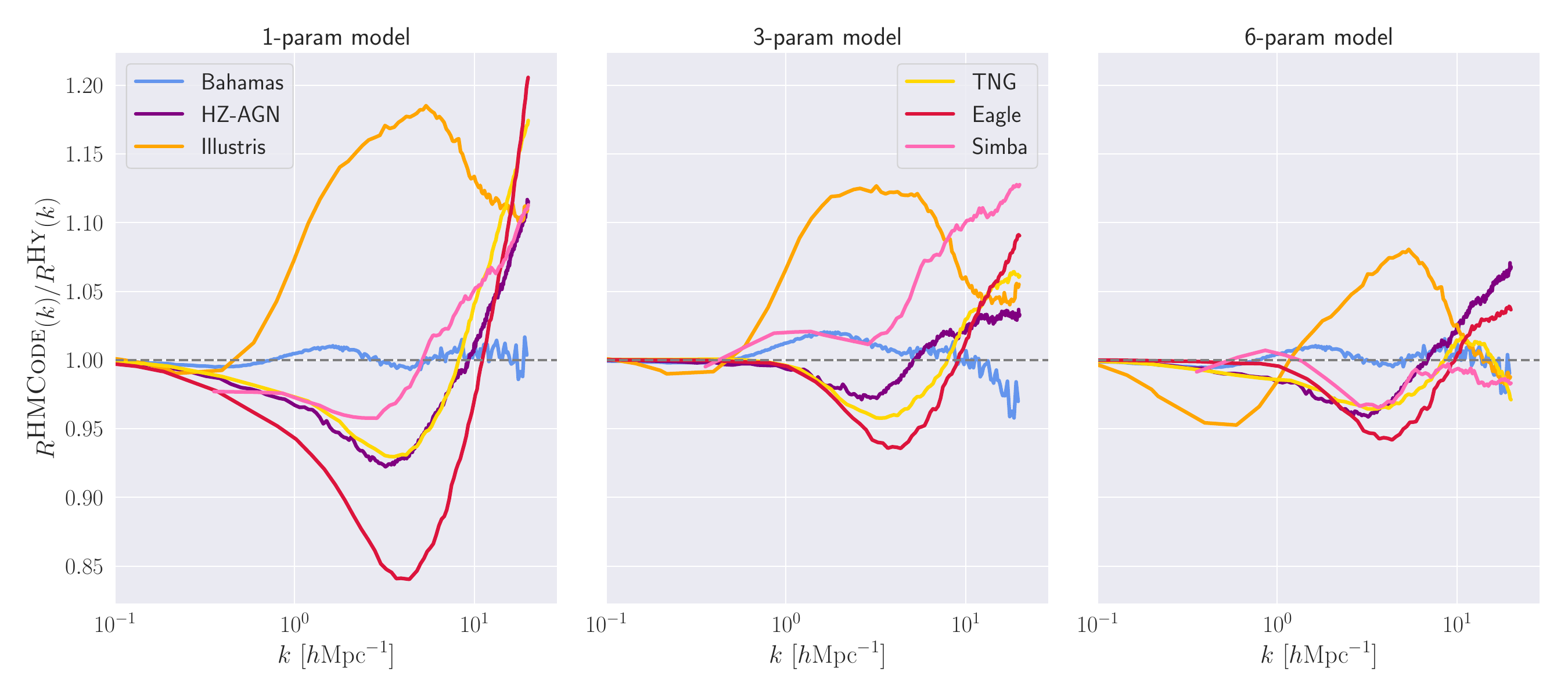}
  \vspace*{-0.25cm}
  \caption{Ratios of the best-fitting \HMCode one-, three-, and six-parameter 
    models to the underlying hydro-sims baryonic response function $R$ at redshift $z=0$.
    We see that \HMCode tends to fit the data better, as given by the smaller
    amplitude, as we increase the generality of the model, which is to be expected.
    Note that we fitted across all redshifts of each of the hydro-sims for 
    $z \leq 2$, where here we are just plotting the $z=0$ slice. 
    Similar curves are found at higher redshifts.}
  \label{fig:R_HMCode_hydros}
\end{figure*}

While our goodness of fit values provide a valuable insight into how well our
three models fit the simulation data overall, we can look at the ratios for
the \HMCode predictions to the hydro-sims results to see how our fit changes
with scales. Figure~\ref{fig:R_HMCode_hydros} plots the ratio of the best-fit \HMCode
model to the hydro-sims baryonic response function for our three $n$-parameter
models as a function of scale at redshift $z=0$. 
We see that the more general three- and six-parameter models are
better able to match the hydro-sims than the one-parameter model, which is to be 
expected from more general models, and is shown by a reduced amplitude in the
relative differences. We also see the extreme nature of the \Illustris 
simulation, where the six-parameter model can only poorly match its feedback. 

\subsection{Constructing the envelope}

Now that we know how well each of our feedback models match our suit of
hydro-sims, we can combine these to form our `envelope function' $\Delta R(k, z)$
introduced in Equation~\ref{eqn:max_deviation}. $\Delta R$ encodes the maximum
deviation of each baryon feedback model to all hydro-sims as a function of
wavenumber and redshift. This acts as a standard deviation to our theoretical
error as each baryon feedback model has an inherent uncertainty and we have 
quantified this with the envelope of the residual discrepancies between the
hydro-sims and the best-fitting $R(k, z)$ for each simulation.

\begin{figure*}
  \includegraphics[width=2\columnwidth]{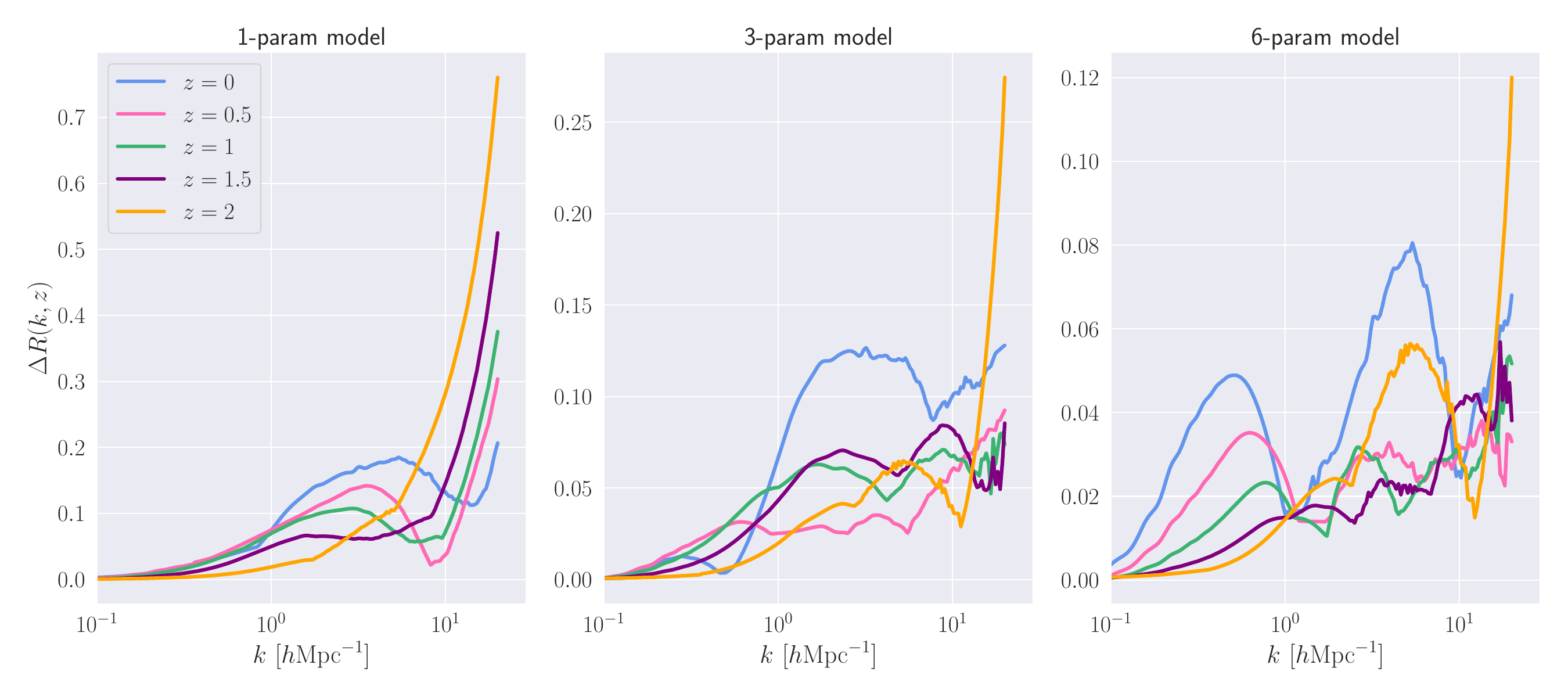}
  \vspace*{-0.25cm}
  \caption{Our envelope function $\Delta R(k, z)$ encoding the maximum deviation of
    our three \HMCode models to the hydro-sims as a function of wavenumber $k$
    plotted for select redshifts. We see that the amplitude of our envelope
    decreases as we go from one- to three- to six-parameters (as noted by the
    decreasing values in each of the panel's individual $y$-axes), and so we assign less
    theoretical uncertainties to those models that better match the data. The
    apparent noisy behaviour of these curves is due to numerical noise in the
    hydro-sims, but also from our choice of envelope function, which uses
    the absolute value of the error ratio. 
    }
  \label{fig:Pk_envelope}
\end{figure*}

We plot $\Delta R(k, z)$ in Figure~\ref{fig:Pk_envelope}, which shows our envelope function as a
function of wavenumber $k$ for select redshift values. We see that, in general,
the errors increase with $k$ which corroborates our understanding that our
theoretical models do worse the smaller scales we probe. We also see that
higher redshifts tend to produce a better fit across all of the models which can
be understood through Figure~\ref{fig:My_Pk_ratio}: we see that at higher
redshifts, the suppression due to baryons is reduced, smoothing our response 
function $R$, and also moving the suppression to smaller wavenumbers. This acts
to allow our baryonic feedback models to better match the hydro-sims, though
our errors do increase to very large values on at high $k$ at high redshifts.

We note that the six parameter model has a distinctive double-hump feature,
which is a direct result of the poor fit to \Illustris on scales
$10^{-1} h \textrm{Mpc}^{-1} < k < 1 h \textrm{Mpc}^{-1}$, as shown in 
Figure~\ref{fig:R_HMCode_hydros}. One could remove \Illustris when constructing
the maximum deviation envelope, and such this double-hump would no longer appear. 

\subsection[Constructing the ell-space theoretical uncertainty covariance matrix]{Constructing the $\bm \ell$-space theoretical uncertainty covariance matrix}

\begin{figure}
  \includegraphics[width=\columnwidth,trim={0.0cm 0.0cm 2.0cm 0.0cm},clip]{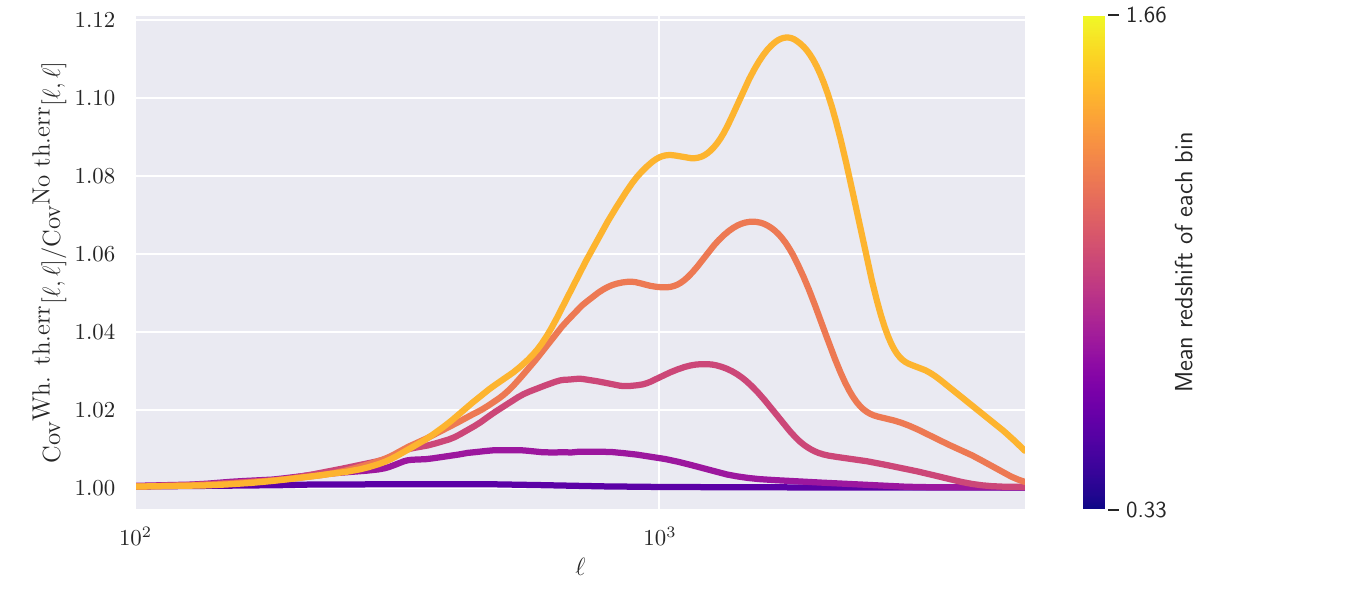}
  
  \vspace*{-0.25cm}

  \caption{Ratio of the diagonals of auto-spectra elements of our $\Cl$ covariance matrix with 
    our additional theoretical error to that without theoretical error. Here,
    we see that the amplitude of the additional theoretical error strongly
    depends on the redshift bins of the auto-spectra considered, with higher redshift bins
    containing more theoretical error than closer bins. We note that there is
    significant support on the off-diagonals for our theoretical covariance, and
    thus the total covariance with theoretical error is much larger than just
    the ratios showed here. 
    }

  \label{fig:Cl_cov_diag}
\end{figure}

Our numerical envelope in $(k, z)$-space can then be doubly integrated (Equation~\ref{eqn:cov_dbl_integral})
to give us our $\ell$-space covariance matrix. Figure~\ref{fig:Cl_cov_diag} plots
the ratio of the block diagonals of the covariance matrix with our additional
theoretical error to that without theoretical error. We see that for $\ell$ modes
below $\ell \simeq 200$ there is no effect of our theoretical error covariance,
since these $\ell$ modes are unaffected by baryon feedback physics. Above this,
we see that out theoretical error covariance acts to increase the total covariance,
thus suppressing $\ell$ modes here and down-weighting them in our likelihood
analyses. We note that this is strongly dependent on the combination of spectra
considered, with low redshift bins having a smaller error than those at high
redshift (which is a direct consequence of our baryon feedback modelling being
less accurate at high redshift). We see that, because we are considering the
covariances of the auto-spectra only, that the ratios tend to unity at
high-$\ell$ which is due to the inclusion of shape noise (which dominates the
auto-spectra for large $\ell$ modes) in the overall covariance matrix.

\begin{figure}
  \includegraphics[width=\columnwidth,trim={0.0cm 0.0cm 0.0cm 0.0cm},clip]{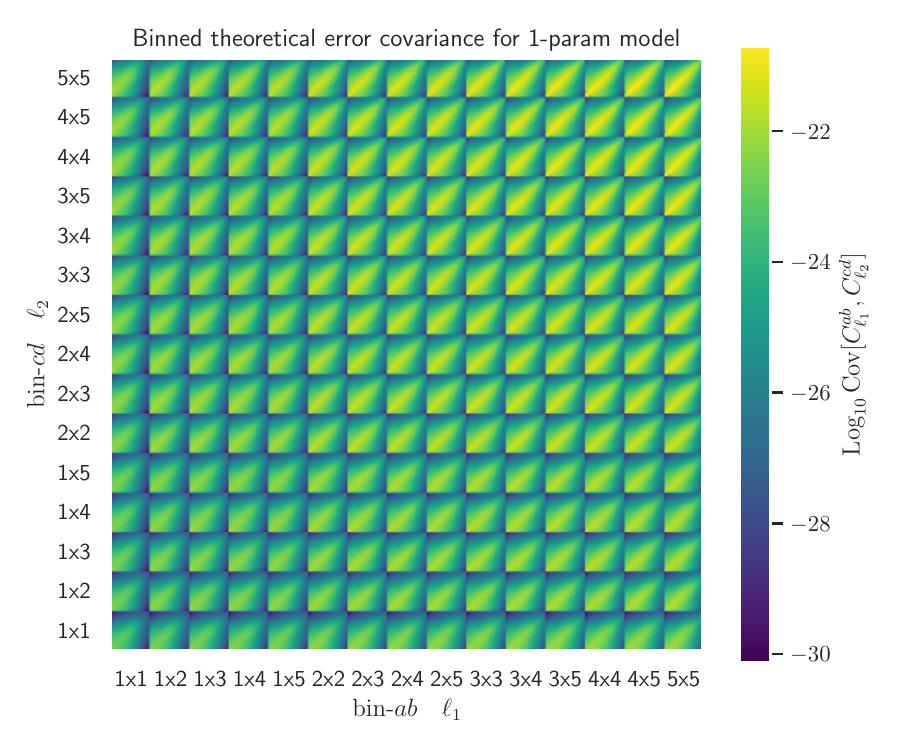}
  
  \vspace*{-0.25cm}

  \caption{Plot of the binned theoretical error matrix $\mathbfss{C}^{\textsc{e}}$
    for the one-parameter model across all fifteen redshift bin combinations and
    twenty five binned band-powers. Here, we see that the amplitude of our
    theoretical error matrix generally grows as we consider higher redshift bins
    (e.g. the 5x5$\times$5x5 sub-block is significantly brighter, and thus
    of larger magnitude, than the 1x1$\times$1x1 sub-block). We also see that
    for each sub-block, the theoretical error is greatest along the 
    $\ell_1 = \ell_2$ diagonal, though there is significant support along the
    off-diagonal terms, which serves to increase the overall effect of our
    theoretical uncertainty modelling.
     }

  \label{fig:Cl_cov_mat}
\end{figure}

We plot the total theoretical uncertainty matrix, $\mathbfss{C}^{\textsc{e}}$
in Figure~\ref{fig:Cl_cov_mat}. Here, we see that the amplitude of the full
matrix reflects what is shown in the diagonals only of Figure~\ref{fig:Cl_cov_diag},
that the amplitude of the theoretical uncertainties generally grow as we consider 
further away redshift bin combinations (going from bottom-left to top-right).
We also see that while each sub-block of the full matrix peaks along its diagonal,
it has significant support with comparable values for many off-diagonal entries,
which acts to boost the effects of the theoretical uncertainties over what
the diagonal values can provide.

\subsection{Parameter constraints and biases}

Using our three different baryon feedback models, we can run our MCMC pipeline
to estimate the biases in our cosmological parameters, the total matter density $\Omegam$
and the lensing amplitude $\Seight$, with and without our additional theoretical
error. We take each hydrodynamical simulations' baryon feedback response function
as the ground truth input values into our cosmic shear pipeline 
(Equation~\ref{eqn:baryon_response_func}).
We run our analyses for each baryon feedback model against each hydro-sim,
with and without the additional theoretical error to
determine the biases on the cosmological parameters due to the effects of
baryon feedback.

\subsubsection{Binary scale-cuts}
\begin{figure}
  \includegraphics[width=\columnwidth]{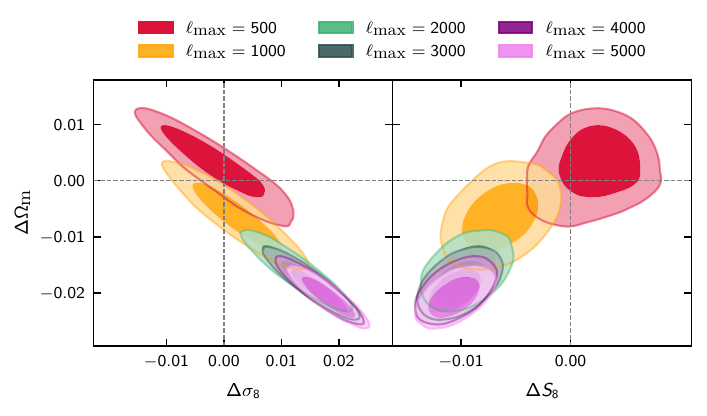}
  
  \vspace*{-0.25cm}

  \caption{2D $\Omegam$-$\sigmaeight$ and $\Omegam$-$\Seight$ offset contours
    for the one-parameter baryon feedback model for the \Simba hydro-sim for
    a range of maximum $\ell$-modes considered in the analysis. As we increase
    $\lmax$, we are increasing the number of modes that are
    contaminated by baryon feedback, and thus bias our cosmological parameters
    away from the fiducial value. It is interesting to note that that even for
    the $\lmax = 1000$ analysis, which is a considerably conservative scale-cut
    even for Stage-III surveys (e.g. HSC-Y3, \citealt{Dalal:2023olq}), our results
    are biased by more than $2\sigma$ in all cosmological parameters for our
    Stage-IV-like survey when using \Simba as the ground-truth.
    }

  \label{fig:simba_scale_cuts}
\end{figure}

We first look at the value of the biases in $\Omegam$ and $\Seight$
as a function of maximum multipole 
when using a traditional binary scale-cuts approach. Figure~\ref{fig:simba_scale_cuts}
plots the $\Omegam$, $\sigmaeight$, and $\Seight$ offset contours for the Simba hydro-sim
when analysed using the one-parameter model for a range of maximum $\ell$-modes
allowed in the analysis. We see that only the $\lmax = 500$ analysis produces
results that are consistent with the input parameters to within $1\sigma$, with
even the $\lmax = 1000$ analysis producing biased results. Each subsequent
analysis where we increase the maximum $\ell$-mode serves to both increase the
raw bias in the cosmological parameters (since the high $\ell$-modes capture 
more of the impact from baryon feedback) and reduce the area of the contours (since
we are now including the more constraining higher $\ell$-modes), which vastly
increases the relative bias in the cosmological parameters.

Hence, if we were to introduce a binary scale-cut for our data that keeps the
$\Delta \Omegam - \Delta \Seight$ to be consistent within $1\sigma$, a
cut slightly larger than $\lmax = 500$ might be made for our idealised
Stage-IV like cosmic shear survey. However, to avoid making these analysis 
decisions ourselves, which have little physical motivation behind them and
apply simultaneously to all hydro-sims in our ensemble, we can turn to our
theoretical error modelling instead of making binary scale-cuts. 

\subsubsection{Including a theoretical error covariance}

Tables~\ref{tab:Omega_m_biases} and~\ref{tab:S8_biases} summarises our results
for the biases in $\Omegam$ and $\Seight$, respectively. Here, we present the
amplitude of the bias, that is the difference in the mean of our MCMC chains
to the fiducial value, and the relative bias, which is the difference divided by
the reported standard deviation from the MCMC chains for each parameter, with
and without the effects of our additional theoretical error. The baryonic
feedback parameters for each model are marginalised over when we present
results for $\Omegam$, $\sigmaeight$ and $\Seight$.

\subsubsection{One-parameter model}

In general, the one-parameter model without additional theoretical 
error produces a significant biases in the recovered parameters, with at least a
3$\sigma$ offset in either $\Omegam$ or $\Seight$ for all hydro-sims excluding
\Bahamas (recall that the one-parameter model was constructed by fitting to
\Bahamas data only). The results of the very significant biases in the remaining
five hydro-sims shows that the one-parameter model of baryon feedback for a
\textit{Euclid}-like  survey up to $\lmax = 5\,000$ is completely impractical if
we wish to recover unbiased results from the analyses of cosmic shear data. 

When we introduce our additional theoretical error into the analysis, we see
that biases in both cosmological parameters fall dramatically across all 
hydro-sims for our one-parameter model. We see a reduction in both the raw
offset values, and a significant decrease in the relative bias in terms of each
parameter's standard deviation. This demonstrates our theoretical uncertainties
modelling is correctly identifying the scales in which baryonic feedback are
not correctly modelled by the one-parameter model, down-weighting them in
our analyses, and thus resulting in significantly less biased cosmological
parameter constraints. However, we still see significant ($>\!1\sigma$)
biases in many hydro-sims even when using theoretical error. This is due to the
one-parameter model being inadequate when considering a \textit{Euclid}-like
weak lensing survey.

\subsubsection{Three- and six-parameter models}

As expected, the more general three- and six-parameter models result in smaller
absolute biases, which reflects their ability to better match more general
baryon feedback scenarios, but at the cost of decreased precision due to the
need to marginalise over more parameters. Hence, we see that the relative
biases show a strong decrease when going to our many-parameter versions, which
is an effect of decreased absolute bias and increased uncertainties.

\begin{figure}
  \includegraphics[width=\columnwidth]{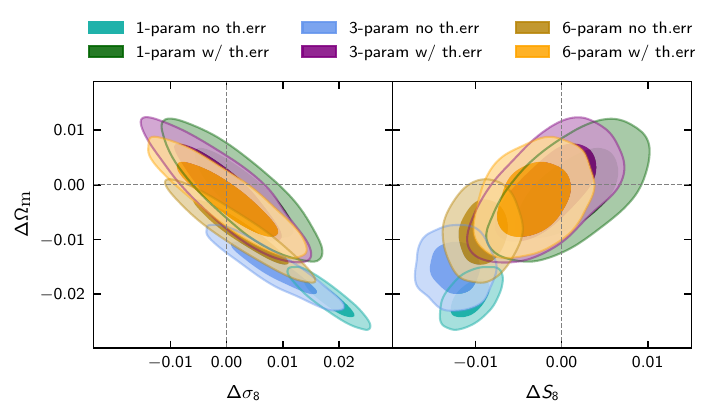}
  
  \vspace*{-0.25cm}

  \caption{2D $\Omegam$-$\sigmaeight$ and $\Omegam$-$\Seight$ offset contours
    for the three baryon feedback models with and without our theoretical
    error for the \Simba hydro-sim. We can clearly see that the one-parameter
    model without theoretical error (light grey contour) produces a tight 
    constraint on our cosmological parameters that is extremely
    far from the correct value. When we add our theoretical error to the data
    covariance (dark green contour), we find that the one-parameter model 
    \textit{is} able to recover the correct values, with an appropriate
    increase in the size of the contour. We see that the three- and 
    six-parameter models, while more flexible, are unable to correctly
    recover unbiased results without the addition of our theoretical
    uncertainties.
    }

  \label{fig:Simba_2D_contour}
\end{figure}

Figure~\ref{fig:Simba_2D_contour} plots the 2D $\Omegam$-$\sigmaeight$ and 
$\Omegam$-$\Seight$ offset contours for our three baryon feedback models with
and without our theoretical error added for the Simba hydro-sim. This highlights
the degeneracies that exist between these parameters, finding the usual `lensing
banana' that is the natural degeneracy between $\Omegam$ and $\sigmaeight$. 
These contours gradually move to the origin as we increase the number of
parameters in the baryon feedback model, and increase in size as we are
marginalising over more parameters.

\begin{figure}
  \includegraphics[width=\columnwidth]{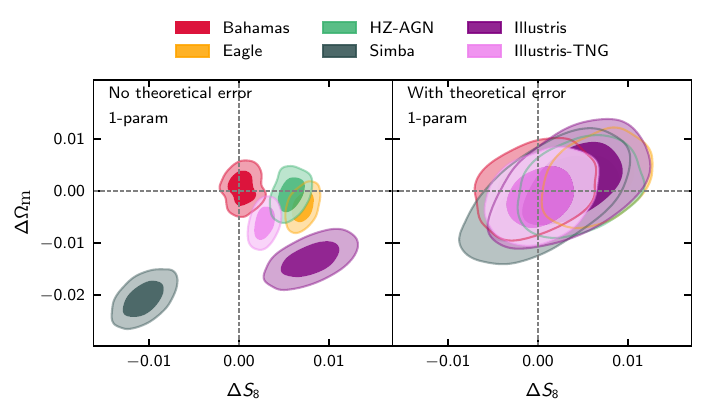}
  
  \vspace*{-0.25cm}

  \caption{2D $\Omegam$-$\Seight$ offset contours for the one-parameter
    baryon feedback model without and with the addition of our theoretical
    error for our range of hydro-sims considered. We can see that without
    theoretical error, the one-parameter model produces a large degree of scatter
    for all hydro-sims, with only \Bahamas being consistent with the input
    cosmology. 
    With the introduction of theoretical error, we see that the contours 
    increase and all except \Eagle become unbiased at the $2 \sigma$ level.
    }

  \label{fig:2D_one_param}
\end{figure}

Figure~\ref{fig:2D_one_param} plots the 2D $\Omegam$-$\sigmaeight$ offset 
contours for the one-parameter model for our suite of hydro-sims for the case of
with and without our theoretical error. We see that without the theoretical
error the recovered contours are highly constraining, which is a result of
marginalising over a single baryon feedback parameter only, with a large degree
of scatter between the hydro-sims. It is interesting to note that this scatter
appears to be roughly along the degeneracy direction, though it appears random
which quadrant each hydro-sim falls into.

\begin{figure}
  \includegraphics[width=\columnwidth]{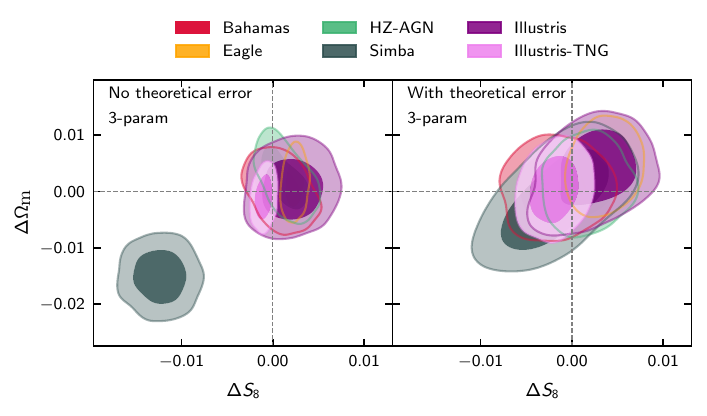}
  
  \vspace*{-0.25cm}

  \caption{2D $\Omegam$-$\Seight$ offset contours for the three-parameter
    baryon feedback model without and with the addition of our theoretical
    error for our range of hydro-sims considered. When compared to the less
    flexible one-parameter model, we see that the three parameter model without
    additional theoretical error is better able to recover the input cosmology
    for a wider range of simulations, though \Eagle, \TNG, and \Simba remain
    biased to at least the $2 \sigma$ level. We note the general increase in
    the contour's area going from the one- to three-parameter model which is
    associated with marginalising over more baryon feedback parameters.
    We see that with theoretical error, all contours are unbiased at the 
    $2\sigma$ level. 
    }

  \label{fig:2D_three_param}
\end{figure}

\begin{figure}
  \includegraphics[width=\columnwidth]{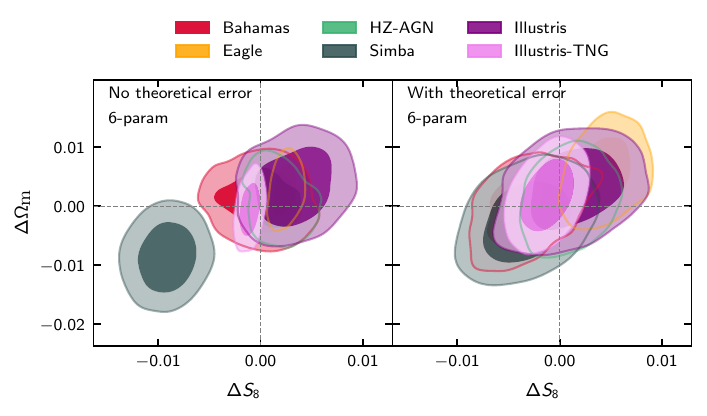}
  
  \vspace*{-0.25cm}

  \caption{2D $\Omegam$-$\Seight$ offset contours for the one-parameter
    baryon feedback model without and with the addition of our theoretical
    error for our range of hydro-sims considered. Again, we see an extension
    of the general trend in the contours without additional theoretical error:
    an increase in the contours with them moving towards the origin. We see that
    even with a six-parameter model, it is not sufficient to recover unbiased
    constraints from the \Simba simulation.
    }

  \label{fig:2D_six_param}
\end{figure}

\begin{table*}
  \caption{Table summarising our results for the bias in $\Omegam$ as found from
    our MCMC analyses using our suite of six hydro-sims and three baryon feedback
    models, without and with our additional theoretical error modelling added
    to the covariance matrix when going to $\lmax = 5000$.
    We see that without theoretical error, the 
    one-parameter model generally produces significant biases ($> \! 3 \sigma$) 
    from the true value, with only \Bahamas giving small biases in $\Omegam$
    and $\Seight$.
    These biases reduce to less than $1 \sigma$ with the inclusion of our 
    theoretical error in the one-parameter model, highlighting the effectiveness
    of our theoretical error modelling. We see that the three- and six-parameter
    result in less bias in $\Omegam$, though some hydro-sims still give large
    biases for the six-parameter model (i.e. \Simba and \Eagle). The inclusion
    of our theoretical error covariance into the three- and six-parameter models
    further reduces the bias.
    }
  \label{tab:Omega_m_biases}
  
  \begin{tabular}{lrrrr}
    \hline
    Hydro-sim and model
    & \quad $100 \Delta \Omegam$ No th. err. &  \quad $\Delta \Omegam / \sigma_{\Omegam}$  No th. err. 
    & \quad $100 \Delta \Omegam$ With th. err. & \quad  $\Delta \Omegam / \sigma_{\Omegam}$ With th. err. \\

    \hline
    
    \Bahamas 1-param & 0.057 & 0.258 & 0.038 & 0.098 \\ 
    \Bahamas 3-param & 0.012 & 0.038 & 0.049 & 0.127 \\ 
    \Bahamas 6-param & 0.100 & 0.276 & -0.085 & -0.206 \\[0.5em]

    \Eagle 1-param & -0.306 & -1.535 & 0.224 & 0.587 \\
    \Eagle 3-param & 0.149 & 0.520 & 0.433 & 1.186 \\
    \Eagle 6-param & 0.274 & 0.962 & 0.573 & 1.459 \\[0.5em]

    \HZAGN 1-param & -0.057 & -0.255 & 0.079 & 0.199 \\
    \HZAGN 3-param & 0.224 & 0.631 & 0.123 & 0.331 \\
    \HZAGN 6-param & 0.109 & 0.325 & 0.105 & 0.269 \\[0.5em]

    \Illustris 1-param & -1.317 & -5.690 & 0.128 & 0.249 \\
    \Illustris 3-param & 0.048 & 0.127 & 0.373 & 0.828 \\
    \Illustris 6-param & 0.328 & 0.761 & 0.295 & 0.652 \\[0.5em]

    \Simba 1-param & -2.071 & -8.828 & -0.158 & -0.293 \\
    \Simba 3-param & -1.513 & -4.723 & -0.155 & -0.276 \\
    \Simba 6-param & -0.858 & -2.206 & -0.257 & -0.567 \\[0.5em]

    \TNG 1-param & -0.613 & -2.885 & -0.107 & -0.277 \\
    \TNG 3-param & -0.110 & -0.424 & 0.035 & 0.091 \\
    \TNG 6-param & -0.055 & -0.185 & 0.181 & 0.453 \\[0.5em]
    
    \hline
  \end{tabular}
\end{table*}  

\begin{table*}
  \caption{Table similar to Table~\ref{tab:Omega_m_biases} but now for the bias in $\Seight$ as found from
    our MCMC analyses. We see that without theoretical error, the 
    one-parameter model results in at least a 3-\!$\sigma$ bias from the true value
    for all hydro-sims except \Bahamas, which reduces when going to a multi-parameter model.
    Our theoretical error is able to significantly reduce the bias in the 
    one-parameter model, though we see a multi-parameter model is essential to
    ensure unbiased constraints across all hydro-sims.
    }
  \label{tab:S8_biases}
  
  \begin{tabular}{lrrrr}
    \hline
    Hydro-sim and model 
    & \quad $100 \Delta \Seight$ No th. err. & \quad $\Delta \Seight / \sigma_{\Seight}$  No th. err. 
    & \quad $100 \Delta \Seight$ With th. err. & \quad $\Delta \Seight / \sigma_{\Seight}$ With th. err. \\

    \hline
    
    \Bahamas 1-param & 0.038 & 0.386 & -0.028 & -0.101 \\
    \Bahamas 3-param & 0.083 & 0.492 & -0.210 & -0.819 \\
    \Bahamas 6-param & 0.010 & 0.042 & -0.360 & -1.468 \\[0.5em]

    \Eagle 1-param & 0.709 & 8.802 & 0.660 & 2.579 \\
    \Eagle 3-param & 0.240 & 3.742 & 0.360 & 2.039 \\
    \Eagle 6-param & 0.240 & 3.207 & 0.441 & 2.443 \\[0.5em]

    \HZAGN 1-param & 0.574 & 6.002 & 0.464 & 1.616 \\
    \HZAGN 3-param & 0.137 & 0.865 & 0.158 & 0.759 \\
    \HZAGN 6-param & 0.165 & 1.070 & 0.089 & 0.448 \\[0.5em]

    \Illustris 1-param & 0.796 & 3.710 & 0.377 & 1.024 \\
    \Illustris 3-param & 0.208 & 0.950 & 0.253 & 0.867 \\
    \Illustris 6-param & 0.325 & 1.289 & 0.135 & 0.435 \\[0.5em]

    \Simba 1-param & -1.059 & -7.096 & 0.081 & 0.207 \\
    \Simba 3-param & -1.235 & -6.469 & -0.180 & -0.480 \\
    \Simba 6-param & -0.916 & -4.903 & -0.328 & -1.156 \\[0.5em]

    \TNG 1-param & 0.281 & 3.701 & 0.033 & 0.130 \\
    \TNG 3-param & -0.100 & -1.602 & -0.193 & -1.083 \\
    \TNG 6-param & -0.106 & -1.628 & -0.122 & -0.715 \\[0.5em]

    \hline
  \end{tabular}
\end{table*}

\subsubsection{Dependency on $\sigma_k$ and $\sigma_z$}
\label{sec:sigma_kz}

In our propagated covariance matrix (Equation~\ref{eqn:k_space_cov}) we introduced
two coupling scales, $\sigma_k$ and $\sigma_z$, which were both set at $0.25$
in our fiducial analyses. Here, we investigate the effects of changing their
values, and the effects it has on parameter constraints. We note that in our
fiducial analyses, the baryon feedback physics of the \Simba simulation is strong
enough that even the six-parameter model of baryon feedback produces significantly
biased cosmological parameter constraints (Figure~\ref{fig:2D_six_param}).
Thus, \Simba is a good testing ground to see how the contours react to the
changing of these values.

In general, as $\sigma_k$ and $\sigma_z$ increase we are increasing the number 
of wavenumber and redshift modes that contribute to a given $(\ell_1, \, \ell_2)$
pair, respectively. This increases the size our $\ell$-space theoretical error
covariance, further suppressing the high-$\ell$ modes that are contaminated
by baryonic feedback. However, in the limit that $\sigma_k$ and $\sigma_z$ 
tend to infinity, then both Gaussian terms tend to unity and the double integral
becomes factorisable in $(k, \, k')$-space and thus is equivalent to some
rank-1 update of the form $\mathbfit{x} \, \mathbfit{x}^{\textsc{t}}$, 
where $\mathbfit{x}$ is some vector in $\ell$-space. Since this is a rank-1
addition to our data covariance matrix, this is equivalent to marginalisation
over a single parameter. However, since we know that single parameter models
for baryonic feedback do not correctly model the wide range of behaviour of
baryonic feedback effects seen in the hydro-sims, it would be incorrect
to take this limit for the values of $\sigma_k$ and $\sigma_z$.

The regime in which our theoretical error covariance reduces to a rank-1 matrix
will be determined by the maximum $k$-space difference between different
$\ell$-modes. For example, Figure~\ref{fig:Cl_deriv_cumsum} shows that there is
an approximate two decade difference in log-$k$ between our smallest and
largest $\ell$-mode. Thus, if one were to choose a value for $\sigma_k$ that
easily encompassed these values within one standard deviation, then we would
expect our theoretical error covariance to collapse to the rank-1 limit.
Since our values chosen for $\sigma_k$ are much less than two decades, we 
expect that our values to not fall within this limit.

\begin{figure}
  \includegraphics[width=\columnwidth]{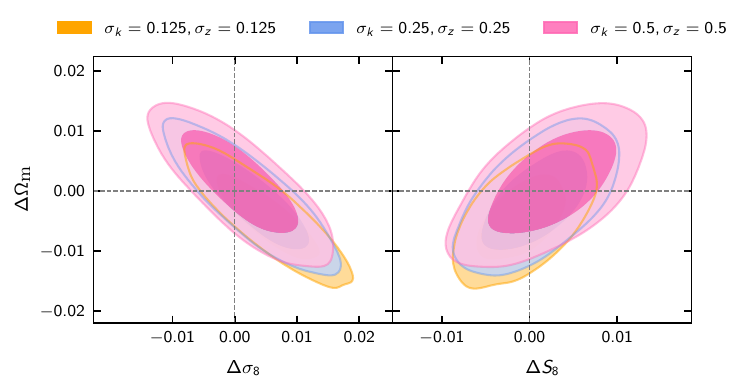}
  
  \vspace*{-0.25cm}

  \caption{2D $\Omegam$-$\sigmaeight$ and $\Omegam$-$\Seight$ contours for 
  the Simba hydro-sim using the one-parameter model with theoretical error
  where we are changing the values of $\sigma_k$ and $\sigma_z$. Here, we see
  that as we increase their values, the effects of the theoretical error
  covariance become larger - increasing the size and decreasing the biases 
  in the contours.
  }

  \label{fig:2D_sigma_kz}
\end{figure}

In Figure \ref{fig:2D_sigma_kz}, we plot the 2D contours for three different
combinations of $\sigma_k$ and $\sigma_z$. Larger values of
these parameters allows for longer-scale correlations in wavenumber and redshift,
and thus produce a larger amplitude for the theoretical error covariance
matrix. Figure~\ref{fig:2D_sigma_kz} shows that these larger values suppress
high-$\ell$ modes more than for smaller values, which is seen by the larger
contours and a reduction in the bias from baryon feedback. While on the range
of values that we tested larger values increase the effectiveness of the
theoretical error covariance, larger values also increases the correlation
length between different wavenumbers and redshifts in the covariance matrix
which decreases the adaptability of our theoretical error covariance to match
more general models of baryon feedback physics. In the limit that
these values go to infinity, we would be saying that all wavenumbers and redshifts
are $100\,\%$ correlated, which is unphysical since we know that baryonic effects
in the power spectrum are quite local at high $k$. Equally, in the limit that
these values go to zero, we would be maximally destroying information through
the covariance matrix by assuming that each wavenumber and redshift has an
independent error which are uncorrelated between similar modes. Ultimately,
there is less than a $1\sigma$ shift in contours plotted in 
Figure~\ref{fig:2D_sigma_kz}, and so our implementation of the theoretical
error approach is broadly insensitive to the values of $\sigma_{k}$ and
$\sigma_{z}$.

We investigate the effects of changing these coupling parameters in our three-
and six-parameter models in Appendix~\ref{app:sigma_kz}, which shows a similar
trend to the one-parameter model in that larger values produce increased 
contours with less baryonic bias in them.

\subsubsection{Growth-structure split}

The information in the cosmic shear power spectrum (Equation~\ref{eqn:cosmic_shear_powspec})
comes from two pieces: the geometrical factors of the lensing kernels, and
in the matter power spectrum, both depending on the matter density $\Omegam$.
It can be advantageous to decouple this matter
density in two: a factor that describes the geometry of the Universe through
the computation of the comoving distances in the lensing kernel, $\Omega_{\textrm{m geom}}$,
and the growth of structure in the universe that is used in the computation of
the matter power spectrum, $\Omegam$~\citep{Matilla:2017rmu}. This 
growth-structure split of the matter density has been applied to existing cosmic
shear data-sets, including the Dark Energy Survey~\citep{DES:2020iqt,Zhong:2023how}
and the Kilo-Degree Survey~\citep{Ruiz-Zapatero:2021rzl}.

We are motivated to apply this growth-structure split to our approach of
modelling the theoretical uncertainties at the matter power spectrum level. 
By quantifying the errors directly in $P(k)$, and then propagating to the
angular power spectrum, we are hopefully preserving information in the lensing
kernels. Our lensing kernels have no theoretical uncertainties associated with
them, and so we hope to preserve information about $\Omega_{\textrm{m geom}}$
even with our scale cuts present.

\begin{figure}
  \includegraphics[width=\columnwidth]{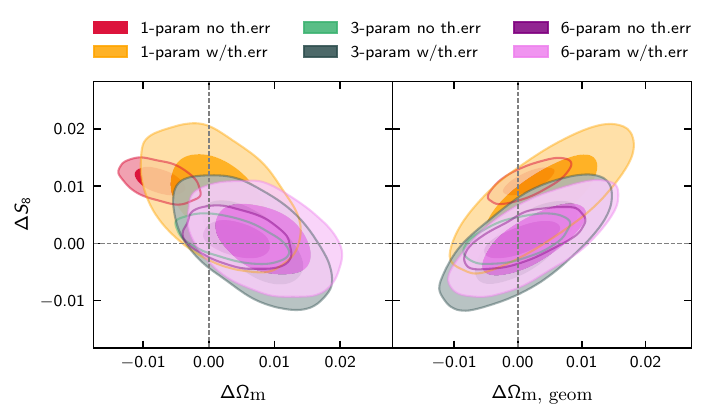}
  
  \vspace*{-0.25cm}

  \caption{2D $\Omegam$-$\Seight$ and $\Omega_{\textrm{m geom}}$-$\Seight$ contours for 
    different baryon feedback models without and with the addition of our theoretical
    error with the \Eagle hydro-sim. We see that without theoretical error, the 
    one-parameter model produces results that are biased in $\Seight$ and $\Omegam$,
    but can identify that the information from the lensing kernels is not affected by
    baryonic feedback, and thus produces unbiased estimates for $\Omega_{\textrm{m geom}}$.
  }

  \label{fig:2D_omegam}
\end{figure}

Thus, we can repeat a sub-set of our MCMC analyses to investigate how the
inclusion of our theoretical error changes the constraints on $\Omegam$ and
$\Omega_{\textrm{m geom}}$. This is presented in Figure~\ref{fig:2D_omegam}. 
Here, we see that without our theoretical uncertainties, 
the inflexible one-parameter model leads to a bias in $\Omegam$
but remains unbiased in the geometrical term $\Omega_{\textrm{m geom}}$. 
This tell us that while our data-vector may still be contaminated with baryonic
feedback from the matter power spectrum, information from the lensing kernels
can still be extracted from cosmic shear data. When we introduce our theoretical
error terms to the covariance matrix, we down-weight baryonic contaminated
modes, and so become consistent in $\Omegam$ and $\Omega_{\textrm{m geom}}$
at the cost of increased contour sizes.

\section{Discussion and conclusions}
\label{sec:diss_and_conc}

We have presented results for parameter constraints using various hydrodynamical
simulations as ground truths analysed with three different baryon feedback models,
with and without an additional theoretical error covariance. We have seen that
constraints on cosmological parameters are significantly biased for a forthcoming
Stage-IV like cosmic shear survey using as low as a maximum $\ell$-mode of
$\lmax = 1000$ when analysing the hydrodynamical simulation's data using
a single parameter for baryon feedback. This is consistent with previous
results in the literature, for example \citet{DES:2020daw} finding a 
harmonic-space scale cut of $\lmax \sim 500$ was needed for the case of
DES-Y3 mock data against a model without baryonic feedback in. 

Hence, we are motivated to more accurately
model the errors associated in the matter power spectrum arising from baryon
feedback over the more traditional binary scale-cuts approach, resulting in
our theoretical error covariance. This theoretical error was then applied to
our model of baryon feedback across our suite of hydrodynamical simulations,
finding that our our parameters biases decreased with smaller absolute and
relative off-sets when going to the same maximum multipole in our analysis. 

We note that some hydro-sims still yield a significant ($> \!\! 1\sigma$) bias
in cosmological parameters when using our theoretical error covariance
for the single-parameter model, which indicates that such a basic model will be
unsuitable for application to Stage-IV cosmic shear survey data.
While the multi-parameter models tended to do better in producing unbiased
results, our theoretical error covariance was still needed to ensure that all
hydro-sim results were unbiased when going to $\lmax = 5000$.

The theoretical error formalism is a general method for quantifying the known
errors of a method to realisations of the data. This method can be applied to a
wide-range of modelling problems within cosmology and astrophysics, where we
have applied it to baryonic feedback within the matter power spectrum. To that
end, the theoretical error is not a fixed quantity. If next generation
hydrodynamical simulations release that feature, for example, improved subgrid models
of physical processes that we know we are currently lacking~\citep{Crain:2023xap},
then it would make sense to replace
the older simulations in our suite with these new releases. This would alter
our error envelope function, improving or reducing our degree of trust in our
numerical baryonic feedback models. Alternatively, as new baryonic feedback
models are developed which are more actuate to the hydro-sims, the need 
for a theoretical error covariance becomes diminished. 

With multi-parameter models of baryonic feedback, the effectiveness of
external priors (that is, information on baryonic feedback not from cosmic shear
alone) become increasingly powerful. In recent years, observational constraints
on baryonic feedback have come from the thermal Sunyaev-Zeldovich effect in
CMB observations~\citep{Troster:2021gsz}, and diffuse x-ray
backgrounds~\citep{Ferreira:2023syi}. With ever increasing precision data
taken on our Universe across the entire electromagnetic spectrum, the power of
external priors and cross-correlation with Stage-IV cosmic shear surveys
for baryonic feedback physics constraints will be immense.

The Holy Grail of
baryonic feedback models will, of course, be a model that can fit all hydro-sims
with no free parameters.\footnote{This is not a requirement on the baryonic
feedback models, per se, but on the requirements on the ensemble of hydrodynamical
simulations to converge to a single prediction for the evolution of $\Phykz$. 
With no scatter in the hydro-sims, deriving a fitting formulae for baryonic
feedback physics becomes trivial.} However, while we are making progress towards this goal,
for example \Bacco~\citep{Arico:2020lhq}, a theoretical error covariance is 
highly valuable until then. We expect the formalism that we have presented
and validated here will be of significant value to forthcoming Stage-IV
weak lensing surveys.

\section*{Acknowledgements}

We acknowledge HPC resources from the IRIS computing consortium.
AH acknowledges support from a Royal Society University Research Fellowship.
For the purpose of open access, the author
has applied a Creative Commons Attribution (CC BY) licence to any Author
Accepted Manuscript version arising from this submission.

\bibliographystyle{mnras}
\bibliography{References/references}

\appendix

\section{Dependency of the covariance on the coupling parameters}
\label{app:sigma_kz}

\begin{figure}
  \includegraphics[width=\columnwidth]{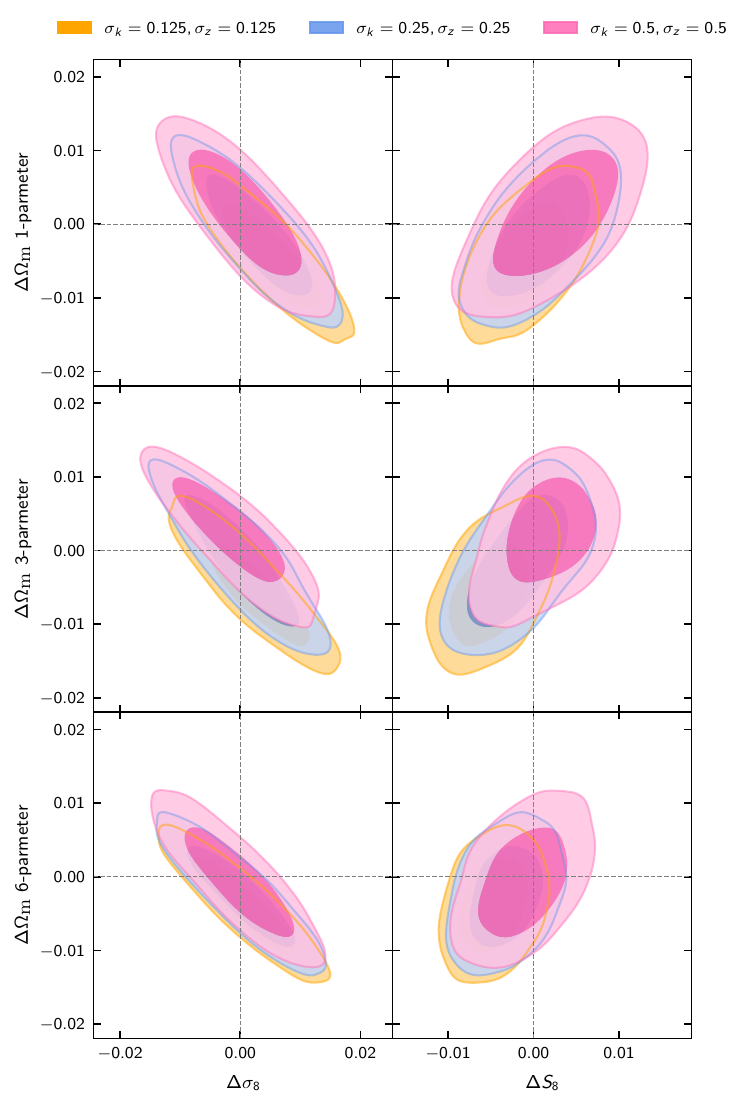}
  
  \vspace*{-0.25cm}

  \caption{$\Omegam$-$\sigmaeight$ and $\Omegam$-$\Seight$ contours for 
  the \Simba hydro-sim using the one-, three-, and six-parameter models (descending rows)
  with theoretical error for different values of $\sigma_k$ and $\sigma_z$ (different coloured contours).
  We see that increasing these values produces a stronger effect for the
  theoretical covariance, increasing the suppression of small-scale modes and
  thus reducing the bias from baryonic feedback.  
  }

  \label{fig:2D_sigma_kz_136param}
\end{figure}

\begin{figure}
  \includegraphics[width=\columnwidth]{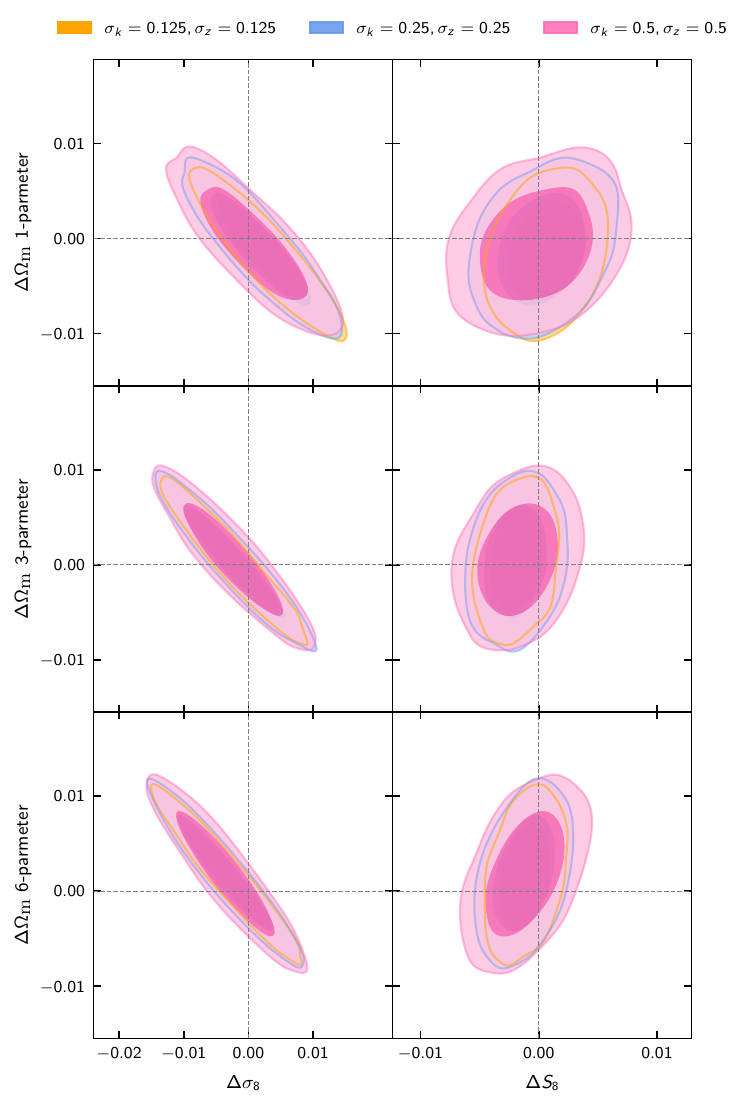}
  
  \vspace*{-0.25cm}

  \caption{Similar plot to Figure~\ref{fig:2D_sigma_kz_136param}, but for the
    \TNG hydro-sim instead of \Simba. Since \TNG can be well fitted by each of our \HMCode
    models with theoretical error, the effect of changing $\sigma_k$ and 
    $\sigma_z$ is just to increase or decrease the size of the contours,
    without effecting the bias in these parameters. 
  }

  \label{fig:2D_sigma_kz_136param_TNG}
\end{figure}

In Section~\ref{sec:sigma_kz} we looked at the one-parameter model and its
dependency on the coupling parameters $\sigma_k$ and $\sigma_z$. Here, we
present a comparison between all three baryon feedback models and investigate
their dependency on the coupling parameters.
Figure~\ref{fig:2D_sigma_kz_136param} uses \Simba as the ground-truth, which
features quite significant baryonic feedback, and
Figure~\ref{fig:2D_sigma_kz_136param_TNG} using \TNG, which has less extreme
feedback. When using \Simba as the ground-truth, we see a that there is a strong
link between the coupling parameter's values and the cosmological parameter 
biases for the one-parameter model, whereas the more general three- and 
six-parameter models are approximately insensitive to $\sigma_k$ and $\sigma_z$
due to their ability to better fit the data (thus have less significant
theoretical uncertainties associated with these models). We also see that,
when using \TNG as the ground-truth, since all three feedback models can 
adequately fit the data, increasing $\sigma_k$ and $\sigma_z$ serves to
slightly broaden the contours only, with no significant effect on the parameter
means.
\clearpage

\bsp	
\label{lastpage}
\end{document}